\documentclass[12pt]{article}	
\usepackage[english]{babel}
\usepackage{amsmath,amssymb,graphicx,enumerate,tabularx,here,xcolor,theorem,mathtools,comment,csquotes,multicol,fancyhdr,lastpage}
\usepackage[backend=biber,bibstyle=ieee,citestyle=numeric,sorting=none,giveninits=true,hyperref=true,maxnames=99,doi=true,arxiv=abs,url=false,eprint=true,backref=false,sortcites=false,abbreviate=true]{biblatex}
\usepackage{bookmark}	
\usepackage{xurl}			
\hypersetup{unicode,bookmarksnumbered=true,hidelinks,final,setpagesize=false,
	bookmarksnumbered=true,bookmarksopen=true,colorlinks=true,linkcolor=blue,citecolor=red}
\usepackage[top=18truemm,bottom=15truemm,left=18truemm,right=18truemm]{geometry}
\numberwithin{equation}{section} 
\title{Unconstrained Lagrangian Formulation for Bosonic Continuous Spin Theory in Flat Spacetime of Arbitrary Dimension}
\author{\\[0em]Hiroyuki Takata\thanks{takata@tspu.edu.ru}
\\[1.5em]
	{\small \it Department of Theoretical Physics, Tomsk State Pedagogical University,}\\
	{\small \it  Kievskaya St.\ 60, 634061, Tomsk,  Russia }
}
\date{\today} 
\begin{document}
\begin{titlepage} 
	\maketitle
	\thispagestyle{empty}
	\begin{abstract}
	  We have discovered two unconstrained forms of free Lagrangian for continuous spin(CS) theory in arbitrary flat spacetime dimension for bosonic case. These Lagrangians, unlike that by Schuster and Toro, do not include delta functions and are conventional. The first form consists of five kinds of totally symmetric helicity fields and one kind of gauge parameter. By introducing auxiliary creation and annihilation operators, each is combined   into a state vector in Fock space, including all ranks one by one. The Lagrangian imposes no constraints, such as trace conditions, on these fields or the gauge parameter field. Additionally, the Lagrangian does not contain higher-order derivative terms. In the limit as CS parameter $\mu$ approaches zero, it naturally reproduces a Lagrangian for helicity fields in higher spin(HS) theory, known as unconstrained quartet formulation. Permitting third-order derivatives, we also obtain the second unconstrained form of Lagrangian that can be written in terms of three kinds of fields, including $\mu$, similar to the formulation by Francia and Sagnotti. Partial gauge fixing and partial use of equations of motion(EOM) on this Lagrangian yield a Fronsdal-like Lagrangian with a single double-traceless field, including $\mu$. By imposing further gauge fixing on the field in the EOM with respect to divergence and trace, we confirm the reproduction of the modified Wigner equations already known in literature.
	\end{abstract}
\end{titlepage}  
{
	\tableofcontents
	\listoftables
	\listoffigures
}
\section{Introduction}\label{intro}

\subsection*{Background}

The Poincare group describes spacetime symmetries in flat spacetime, and its representations give rise to various particles. Wigner {\cite{Wigner:1939cj}} and Bargmann and Wigner {\cite{Bargmann:1948ck}} first studied these representations in detail in four-dimensional spacetime, which was generalized to arbitrary spacetime dimension $D$ {\cite{Bekaert:2006py}}. The representations of the massless little group are classified into two types: helicity representations(=massless representation in standard sense) and continuous spin(CS) representations {\cite{Wigner:1939cj,Bargmann:1948ck,Bekaert:2006py,Iverson:1971hq,Brink:2002zx}}.

Lagrangians describing bosonic and fermionic fields with arbitrary helicity in flat and de Sitter spacetime have been constructed by Fang and Fronsdal {\cite{Fronsdal:1978rb,Fang:1978wz,Fronsdal:1978vb,Fang:1979hq}}. For recent reviews, see, for example, {\cite{Bouatta:2004kk,Bekaert:2004qos,Didenko:2014dwa}}. The Lagrangian for CS fields was first introduced by Schuster and Toro in four-dimensional flat spacetime {\cite{Schuster:2014hca}}(see also{\cite{Rivelles:2014fsa}}). Metsaev extended this Lagrangian to higher dimension, using either a tower of double-traceless fields or two towers of traceless fields {\cite{Metsaev:2016lhs,Metsaev:2018lth}}. For reviews, see, for example {\cite{Bekaert:2017khg}}. The construction of Lagrangians for bosonic and fermionic fields has been extensively studied in {\cite{Zinoviev:2017rnj,Khabarov:2017lth,Bekaert:2015qkt,Metsaev:2017ytk,Metsaev:2017myp,Najafizadeh:2017tin}}.

Unconstrained Lagrangian formulations are formalisms in which no constraints are imposed on the fields or gauge parameters contained in the Lagrangian a priori. These formulations build up on the benefits of Lagrangians. Traditionally, Lagrangians have been used to derive the equations of motion(EOM) that describe the dynamics of a field. However, an ``unconstrained Lagrangian''($\equiv$ unconstrained form of Lagrangian$\equiv$Lagrangian described by unconstrained fields and unconstrained gauge parameters only) is assumed to be able to derive all the necessary information, including constraint conditions for the irreducible representation of a group, in a generalized sense of EOM. Furthermore, unconstrained Lagrangian formulations may also be suitable when considering interactions. For details, please refer to the concluding section \ref{summary}. 

Unconstrained Lagrangian formulations fall into three main categories:
\begin{enumerate}[a{\textup{)}}]
  \item \label{a}{\textbf{Delta function approach:}} This method resolves constraint conditions using delta functions, typically resulting in Lagrangian that contains delta functions but no auxiliary fields.
  
  \item \label{b}{\textbf{Geometric approach:}} This method introduces non-local terms into the Lagrangian to eliminate constraint conditions. It   also includes a method that avoids using non-local terms by introducing a few auxiliary fields, leading to the appearance of higher derivative terms. 
    
  \item \label{c}{\textbf{Auxiliary field approach:}} This method introduces auxiliary fields to achieve an equivalent conventional Lagrangian. Examples include the BRST\footnote{The BRST formulation is sometimes denoted as BRST-BFV formulation for clearer distinction, but in this paper, it will simply be referred to as BRST formulation. It is worth noting that there are also BRST \ formulation that do not incorporate trace constraints and are not fully unconstrained(e.g.{\cite{Alkalaev:2017hvj,Alkalaev:2018bqe,Burdik:2019tzg}}). However, in this paper, the term ``BRST'' will always refer to the fully unconstrained version, including trace constraints.
  }
  formulation and the quartet unconstrained formulation.
\end{enumerate}

All three types have been studied in higher spin(HS) theories (i.e., without CS parameter $\mu$). In fact, the delta function approach \ref{a}) was employed by Segal in {\cite{Segal:2001qq}} and by Najafizadeh in {\cite{Najafizadeh:2018cpu}}. Francia and Sagnotti used the geometric approach \ref{b}), both with non-local terms {\cite{Francia:2002aa,Francia:2002pt}} and with local higher derivative terms {\cite{Francia:2005bu}}. The auxiliary field approach \ref{c}) was applied by Buchbinder et al. in the quartet formulation{\cite{Buchbinder:2007ak,Buchbinder:2008ss}} and by various authors in the BRST formulation {\cite{Pashnev:1998ti,Burdik:2001hj,Buchbinder:2001bs,Buchbinder:2002ry,Bekaert:2003uc,Buchbinder:2004gp,Buchbinder:2005ua,Buchbinder:2006nu}}.

In particular, Buchbinder et al. {\cite{Buchbinder:2007ak}} achieved a Lagrangian for any spin in $\mu = 0$ massless case, by introducing auxiliary fields into formulations in {\cite{Francia:2002aa,Francia:2002pt,Francia:2005bu,Francia:2006hp}}. This Lagrangian was subsequently generalized to the massive case {\cite{Buchbinder:2008ss}}. Here, all fields and gauge parameter are trace-full. For example, in bosonic case, the Lagrangian includes six fields, two of which are Lagrange multipliers. These approaches {\cite{Buchbinder:2007ak,Buchbinder:2008ss}}, quartet unconstrained formulations, are particularly interesting because the resulting Lagrangians can also be derived from the BRST formulated Lagrangian for any spin of massless/massive fields. 

The BRST Lagrangian formulation is a fascinating approach that directly connects spacetime symmetry, specifically the Poincare algebra in flat spacetime, to Lagrangian itself. Furthermore, as pointed out in the literature {\cite{Sagnotti:2003qa}}, it is also suggested to be related to string theory, making it an intriguing subject of study. 

On the other hand, for CS theories (i.e., with $\mu \neq 0$), only the delta function approach \ref{a}) has been well studied for bosonic theories {\cite{Schuster:2014hca}}, fermionic theories {\cite{Bekaert:2015qkt}}, and supersymmetric theories {\cite{Najafizadeh:2021dsm}}. Notably, the geometric \ref{b}) and auxiliary field \ref{c}) approaches have not been explored for CS theories at all, except for a BRST approach discussed in {\cite{Bengtsson:2013vra}} and 'quartet-like' description{\cite{Burdik:2019tzg}}.\footnote{In {\cite{Burdik:2019tzg}}, the authors construct constrained descriptions of EOM and Lagrangian using the BRST-BFV formulation. They then introduce auxiliary fields, including a compensator, to obtain an unconstrained Lagrangian (termed 'quartet-like'). Our Lagrangian is minimal, containing only five fields, while theirs contains 14 fields. Thus, the correspondence between their work and ours is still unclear, which is an interesting point to study.
}
This research gap presents a compelling opportunity to both advance our understanding and expand the toolkit available for Poincare group representations in theoretical particle physics. 

\subsection*{Objectives and Overview(see also Figure \ref{fig2})}

In this paper, we aim to obtain conventional, unconstrained Lagrangian formulation that describes fields following the CS representation. In other word, we aim to generalize the results of {\cite{Buchbinder:2007ak}} and {\cite{Francia:2005bu}} to the case of CS (i.e., non-zero $\mu$). We focus on massless, totally symmetric free bosonic fields in flat spacetime of arbitrary dimension. We will derive Lagrangians that reproduce the Wigner equations for CS field. We pay particular attention to obtaining an unconstrained form without field or gauge parameter constraints, non-local terms, or delta functions. This simplifies the Lagrangian for further treatment and analysis.

We present two novel unconstrained Lagrangian formulations:

\begin{enumerate}
  \item A conventional Lagrangian $\mathcal{L}_5$(\ref{Lag_5f}) with up to second-order derivatives and five fields,\footnote{The number of  continuous spin fields is counted as all ranks of tensor fields are included one by one and collectively regarded as a single field. In the following sections of this paper, the subscript number $i$ attached to the Lagrangian:
  $\mathcal{L}_i, i = 1 \sim 5$, represent the number of fields in that sense.
 
 }
 similar to {\cite{Buchbinder:2007ak}} (categorized as \ref{c}) above).
  
  \item A simpler Lagrangian $\mathcal{L}_3$(\ref{Lag_3f}) with up to third-order derivatives and three fields, similar to {\cite{Francia:2005bu}} (categorized as \ref{b}) above).
\end{enumerate}
Both types of unconstrained Lagrangian formulations are novel contributions of this paper and will be presented in detail in Section \ref{lag_unconst}.

The possibility of obtaining an unconstrained Lagrangian of CS theories in arbitrary dimension using the BRST formulation will be discussed in the concluding section \ref{summary}.

\subsection*{Methodology}

Here, we outline the advantages of the formulation used throughout this paper.
\begin{enumerate}[a{\textup{)}}]
  \item \label{meth_a} {\textbf{Use of HS operators and state vectors in Fock space:}}
  
  The Lagrangians and Wigner equations are written using ``Fock state vector''($\equiv$state vector in Fock space) and HS operators, which act on these state vectors and are defined using creation-annihilation operators and differential operators.\footnote{Equations and Lagrangians written in terms of state vectors in Fock space and HS operators can always be rewritten in terms of tensor fields, which are more familiar to many people. For more details on the correspondence, see, for example, {\cite{Buchbinder:2005ua,Buchbinder:2006nu}}.
  }
  This approach clarifies the relationships among Lagrangians and equations and also allows us to exploit the advantages of HS operators, which form an algebraically closed and tractable structure.
  
  \item \label{meth_b}{\textbf{Implicit inclusion of the divergence operator:}}
  
  The divergence operator $l_1 = a^{\mu} \partial_{\mu}$, which is one of the HS operators,\footnote{See Appendix \ref{convention} for definitions.
  }
  appears only in its extended form that implicitly includes the CS parameter $\mu$. It makes the formal similarity to the case of $\mu = 0$ clear because of no explicit appearance of $l_1$ itself.
  
  \item \label{meth_c}{\textbf{Development of mathematical tools:}}
  
  In the course of our research, we developed several useful mathematical tools, including a ``Fourier-like transformation'' from function space to Fock space, a ``trace generating operator'' written in terms of HS operators, and a ``unitary-like operator'' that bridges different Fock spaces.
  
  \item \label{meth_d}{\textbf{Absence of imaginary numbers:}}
  
  To make it clear that the fields considered in this paper can be restricted to real fields, we ensure that no imaginary numbers appear in the operators used in the Wigner equations and Lagrangians.\footnote{We would like to thank I. L. Buchbinder for his suggestions on this point.
  }
  This can be achieved by a suitable phase rotation.
\end{enumerate}

\subsection*{Relation to Other Studies}

\begin{figure}[h]
	\centering
	\includegraphics[width=1.00\hsize]{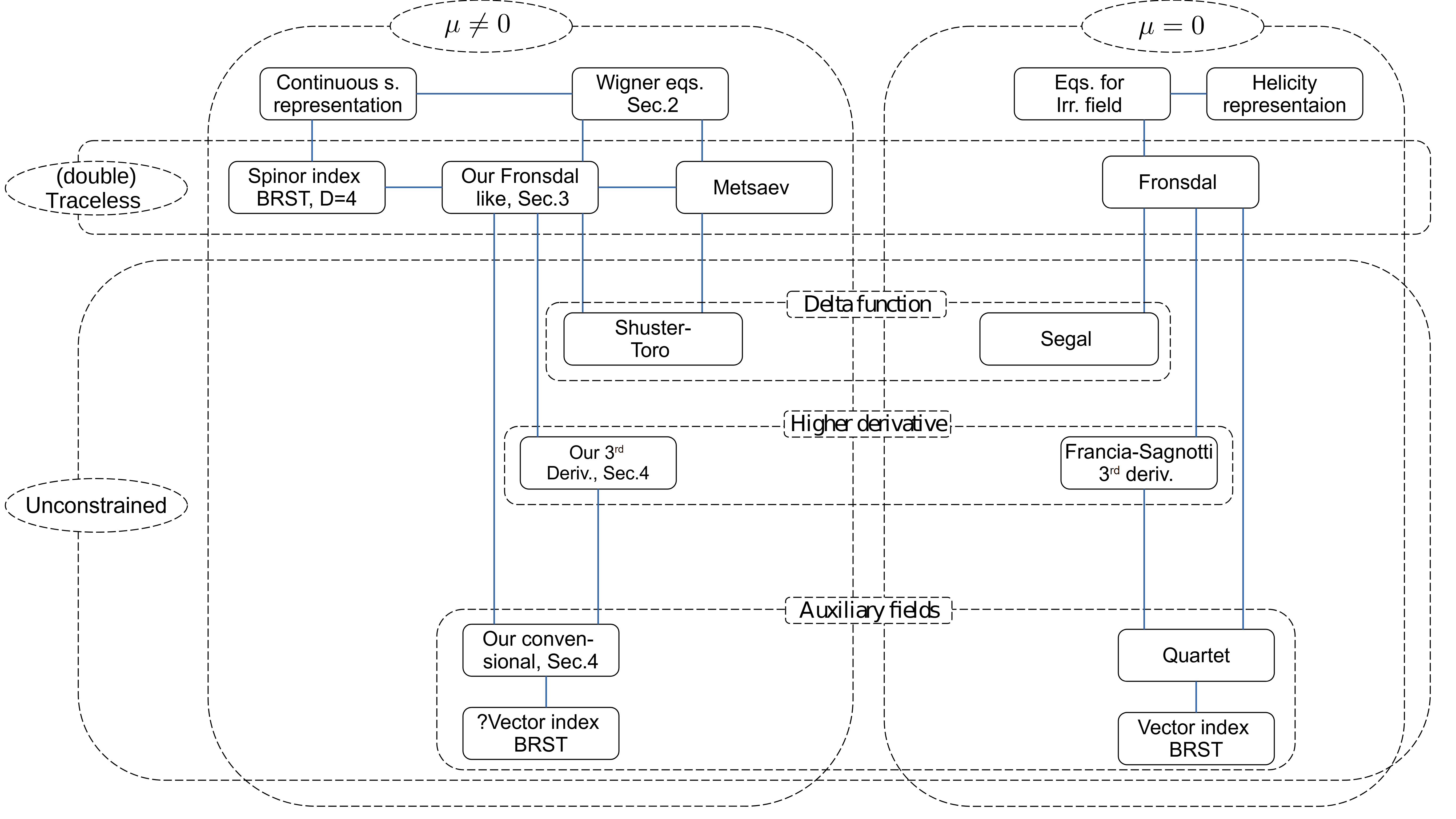}
	\caption[Correlation diagram for various Lagrangians in different studies]{\label{fig1}Correlation diagram illustrating various Lagrangians presented in different studies, including ours. These diagrams depict only Lagrangians relevant to massless totally symmetric bosonic fields in arbitrary flat $D$-dimensional spacetime, excluding case explicitly labeled for $D = 4$. Note that topmost entries are not themselves Lagrangians.}
\end{figure}

Our work is related to several other studies(see Figure \ref{fig1}):
\begin{enumerate}[a{\textup{)}}]
  \item \label{rel_a}{\textbf{(Double) Traceless Formulation by Metsaev}}
  
  Metsaev {\cite{Metsaev:2016lhs,Metsaev:2018lth}} introduced a Lagrangian applicable in any spacetime dimension formulated in terms of either traceless or double-traceless fields and including parameters for mass, CS scale, and the anti-de Sitter curvature. In the special case of massless fields in flat spacetime, this Lagrangian is equivalent to Lagrangian $\mathcal{L}_1$(\ref{Lag_1f}), which can be derived from the unconstrained formulation of our Lagrangians $\mathcal{L}_5$(\ref{Lag_5f}) and $\mathcal{L}_3$(\ref{Lag_3f}) through partial gauge fixing and the partial use of EOM. Additionally, by taking the limit $\mu \rightarrow 0$, Lagrangian $\mathcal{L}_1$(\ref{Lag_1f}) exactly coincides with the Fronsdal Lagrangian \cite{Fronsdal:1978rb} summed over all helicities.
  
  \item \label{rel_b} {\textbf{BRST Formulation with Spinor Index Fields \cite{Buchbinder:2018yoo}}}
  
  The BRST formulation of a Lagrangian using spinor index fields in four-dimensional flat spacetime was previously studied in \cite{Buchbinder:2018yoo}.\footnote{ In this direction, there have been extended studies regarding particle statistics, spacetime dimension, and curvature in \cite{Buchbinder:2020nxn,Buchbinder:2022msd,Buchbinder:2023vkb,Buchbinder:2024jpt,Buchbinder:2024hea}. For formulations with additional spinor variables and in twistor formulation, see also \cite{Buchbinder:2018soq,Buchbinder:2019iwi,Buchbinder:2020ocz,Buchbinder:2021bgv,Buchbinder:2022nvx,Buchbinder:2023qog}.
  }
  By exploiting the relationship between spinor and vector index fields {\cite{Buchbinder:1998qv}}, and through a suitable reformulation of operators in different Fock spaces, this Lagrangian $\mathcal{L}_s$(\ref{Lag_spin}) can be expressed in terms of traceless fields, as shown in Lagrangian $\mathcal{L}_2$(\ref{Lag_2f}). This Lagrangian is also obtainable from the unconstrained formulation of our Lagrangians $\mathcal{L}_5$(\ref{Lag_5f}) and $\mathcal{L}_3$(\ref{Lag_3f}) through partial gauge fixing and the partial use of EOM.
  
  \item  \label{rel_c} {\textbf{Delta Function Formulation by Schuster and
  Toro}}
  
  The relationship between Lagrangian for unconstrained fields by Schuster and Toro{\cite{Schuster:2014hca}} and that of our paper is indirect and established through a model with a constrained field. This indirect connection can be understood similarly to the relationship between the Segal Lagrangian {\cite{Segal:2001qq}} and the quartet Lagrangian {\cite{Buchbinder:2007ak}} and the Lagrangian by Francia and Sagnotti in  {\cite{Francia:2005bu}} in the $\mu = 0$ case.
  
  In the $\mu = 0$ case, the Segal Lagrangian is obtained from the Fronsdal Lagrangian through a field redefinition and solving constraint equations using delta functions, while the quartet Lagrangian {\cite{Buchbinder:2007ak}} and the Lagrangian of {\cite{Francia:2005bu}} are obtained from the Fronsdal Lagrangian by including auxiliary fields(such as Lagrange multiplier and compensator).
  
  In a similar manner, the Lagrangians by Schuster and Toro are obtained from $\mathcal{L}_1$(\ref{Lag_1f}) or the Metsaev type Lagrangian{\cite{Metsaev:2016lhs,Metsaev:2018lth}} through a field redefinition and solving constraint equations using delta functions, while our Lagrangians $\mathcal{L}_5$(\ref{Lag_5f}) and $\mathcal{L}_3$(\ref{Lag_3f}) are obtained from $\mathcal{L}_1$(\ref{Lag_1f}) or the Metsaev type Lagrangian by including auxiliary fields.
  
  Note that: the field redefinition (or modification) mentioned here was first introduced in {\cite{Najafizadeh:2017tin}} for Wigner equations and is used to rewrite the constraints between (double) trace-full conditions and the (double) traceless conditions. In our paper, we reformulate this using
  creation and annihilation operators(see also {\cite{Alkalaev:2017hvj}}). The detailed explanation for why this modification is essential to understand
  the relationship between the previously mentioned models can be found in {\cite{Najafizadeh:2018cpu}}.
  
  \item  \label{rel_d} {\textbf{Quartet Unconstrained Formulation by Buchbinder, Galajinsky, and Krykhtin}}
  
  By setting $\mu = 0$ in our Lagrangian $\mathcal{L}_5$(\ref{Lag_5f}) before applying gauge fixing or using the EOM for auxiliary fields, we recover the unconstrained formulation of Lagrangian for $\mu = 0$ massless case, which is also obtainable from BRST Lagrangian expressed in terms of vector index fields for $\mu = 0$ massless case {\cite{Buchbinder:2007ak}}.
  
  \item \label{rel_e} {\textbf{Local, Higher-Order Derivative Formulation by
  Francia and Sagnotti}}
  
  One of our unconstrained Lagrangian $\mathcal{L}_5$(\ref{Lag_5f}) contains only up to second-order derivatives. However, by using the EOM for auxiliary fields, we can obtain a more concise Lagrangian that incorporates third-order derivatives while maintaining unconstrainedness. This Lagrangian $\mathcal{L}_3$(\ref{Lag_3f}), similar to the formulation in {\cite{Francia:2005bu}}, can still be expressed in terms of only three   fields, even in the presence of $\mu$.
\end{enumerate}
\subsection*{Layout of Paper}

This paper is organized as follows:

In {\textbf{Section \ref{wigner}}}, we first prepare a Fock space with creation and annihilation operators with $D$-dimensional vector indices. We then rewrite the original Wigner equations {\cite{Wigner:1948gr,Bargmann:1948ck}} as equations for state vectors in this Fock space. To achieve this, we introduce a Fourier-like transformation from real auxiliary variables to creation operators. We further rewrite these Wigner equations using HS operators, which have an algebraically closed and manageable structure. Next, we transform these Wigner equations to include the traceless condition. In doing so, we introduce a mathematical tool similar to the exponential operators that generate coherent states, which we call a trace generating operator.

In {\textbf{Section \ref{lag_const}}}, we aim to derive a Lagrangian that reproduces the Wigner equations obtained in Section \ref{wigner}. We begin with a four-dimensional BRST Lagrangian {\cite{Buchbinder:2018yoo}} written in ``spinor Fock space''($\equiv$a Fock space constructed by spinor index creation operators). To extend this to arbitrary $D$-dimensions, our first and main step involves introducing a mathematical tool: a unitary-like operator that maps between spinor Fock space and traceless subspace of ``vector Fock space''($\equiv$a Fock space constructed by vector index creation operators). Using this operator, we obtain a Lagrangian written in terms of vector index fields and subsequently extend it naturally to arbitrary $D$-dimensions. This results in a Lagrangian with a single double-traceless field, which is equivalent to the Metsaev Lagrangian {\cite{Metsaev:2016lhs,Metsaev:2018lth}} and can also be viewed as an extension for CS fields from the Fronsdal Lagrangian {\cite{Fronsdal:1978rb}}. Finally, we perform gauge fixing using a generalized divergence operator and trace operator. This gauge fixing procedure is demonstrably equivalent to the one employing de Donder-type operators {\cite{Metsaev:2016lhs,Metsaev:2008fs}} and the trace operator. We confirm that this procedure successfully reproduces the Wigner equations derived in Section \ref{wigner}. 

In {\textbf{Section \ref{lag_unconst}}}, we begin with the Lagrangian derived in Section \ref{lag_const} and perform equivalent transformations to remove constraints on the fields and the gauge parameter. To ensure that the EOM remain unchanged, we introduce three auxiliary fields and derive a Lagrangian without field constraints. To remove constraint on the gauge parameter, we introduce an auxiliary field called compensator. To this end, we develop a general method for systematically extending Lagrangian and apply it to obtain a conventional Lagrangian without constraints, higher-order derivatives, delta functions, or non-local terms. This Lagrangian has five fields and one gauge parameter, that extends the work in {\cite{Buchbinder:2007ak}} by incorporating the CS parameter $\mu$, while eliminating one auxiliary field. Retaining the compensator field and eliminating two auxiliary fields by their EOM, we arrive at a simpler, unconstrained Lagrangian extending {\cite{Francia:2005bu}} with the inclusion of $\mu$, written in terms of three fields with third-order derivative terms. 

In {\textbf{Section \ref{summary}}}, we conclude paper by summarizing our findings, discussing contributions of our research, and highlighting potential future directions, including discussion about potential directions for BRST formulation of CS Lagrangian.

In {\textbf{Appendix \ref{convention}}}, we provide a summary of the notations
and conventions used in this paper.

In {\textbf{Appendix \ref{fourier}}}, we define and verify the consistency of a Fourier-like transformation between real variables and creation operators. We also summarize simple correspondences between these variables. This enables us to clearly understand the equivalence between expressions in terms of creation-annihilation operators (or HS operators) and state vectors in Fock space, and expressions in terms of more familiar coordinate-space functions.

In {\textbf{Appendix \ref{factorial}}}, we define factorial powers and summarize their relevant properties for binomial polynomials.

In {\textbf{Appendix \ref{traceop}}}, drawing an analogy to coherent state generating operator, we introduce trace generating operator (for arbitrary times) and its inverse. We define these operators using the language of HS operators and the factorial powers. We also explain the derivation method and summarize useful formulas.

In {\textbf{Appendix \ref{m_wigner}}}, we provide a step-by-step explanation of the calculations involved in deriving the modified Wigner equations using HS operators.

In {\textbf{Appendix \ref{casimir}}}, we show that under the modified Wigner equations, Casimir operator acting on Fock state vector has a correct eigenvalue for CS.

In {\textbf{Appendix \ref{alg_wigner}}}, we summarize the key algebraic structure of the Wigner equations relevant to this paper.

In {\textbf{Appendix \ref{compensator}}}, we introduce a general method for introducing a compensator to eliminate trace constraint condition on gauge parameter. We then demonstrate its application to our CS Lagrangian. This is a crucial step in the derivation of the two novel Lagrangians presented in this paper.

In {\textbf{Appendix \ref{eom_equiv}}}, we derive EOM for our unconstrained Lagrangian and demonstrate their equivalence to EOM of Fronsdal-like constrained Lagrangian.

\section{Oscillator Realization and Modification of Wigner Equations}\label{wigner}

This section begins by revisiting the original Wigner equations. We rewrite them in terms of creation and annihilation operators, which allows for a clearer and more rigorous mathematical treatment. We then modify one of these equations to satisfy the traceless condition, which is a crucial step for deriving Lagrangian in Section \ref{lag_const}. This procedure is similar to that in {\cite{Najafizadeh:2017tin}}, but here we express all expressions using creation and annihilation operators that operate on state vectors in Fock space. As a final step, we further rewrite the Wigner equations using HS operators. HS operators have a well-defined algebraic structure, which simplifies algebraic manipulations. In Section \ref{lag_const}, we reproduce these equations from Lagrangian by applying an appropriate gauge fixing.

\subsection{Wigner Equations using Creation and Annihilation Operators}

Wigner equations are fundamental in the study of CS theory, as they are essential tools for studying the irreducibility of CS representations. Originally formulated by Wigner {\cite{Wigner:1948gr,Bargmann:1948ck}} for four-dimensional spacetime, these equations have been extended to arbitrary dimensions $D$ and are expressed as follows{\cite{Bekaert:2006py,Bekaert:2005in}}:
\begin{align}
  p^2 \Psi (p, \xi) & =  0  \label{kg}\\
  (p \cdot \partial_{\xi} - i \mu) \Psi (p, \xi) & =  0  \label{gauge}\\
  (p \cdot \xi) \Psi (p, \xi) & =  0  \label{div}\\
  (\xi^2 - 1) \Psi (p, \xi) & =  0 \quad .  \label{trace}
\end{align}
Here, $\mu$ is the CS parameter which is real and has mass dimension, $p_{\mu}$ represents spacetime momentum, $\xi^{\mu}$ represents $D$-dimensional auxiliary real variables, and $\partial_{\xi} =\frac{\partial}{\partial \xi}$. To realize these equations with oscillators, we introduce auxiliary creation and annihilation operators $a^{\dag \mu}$ and $a^{\mu}$, respectively, along with the vacuum state $|0 \rangle$, defined by $a^{\mu} | 0 \rangle \equiv 0$. These operators satisfy the commutation relation:\footnote{Refer to Appendix\ref{convention} for conventions.
}
\begin{align}
   [a^{\mu}, a^{\dag \nu}] &=  \eta^{\mu \nu} \quad .
\end{align}
Now, we perform a Fourier-like transformation from function $\Psi (\xi)$ of the variables $\xi^{\mu}$ to \ state vector $| \psi \rangle$ in Fock space, generated by the creation operators $a^{\dag \mu}$ as follows:\footnote{Delta function with a general operator $\hat{O}$ is understood as: $ \delta (\hat{O}) =  \int \frac{d s}{2 \pi} e^{s \hat{O}} =  \int  \frac{d s}{2 \pi} \sum_n \frac{(s \hat{O})^n}{n!} $.
}
\begin{equation}
	\begin{alignedat}{2}
		\psi (a^{\dag}) & :=  \int \frac{d^D \xi}{(2 \pi)^{D / 2}} e^{- i    a^{\dag} \cdot \xi} \Psi (\xi)  & &  \\
		\Psi (\xi) & =  (2 \pi)^{D / 2} \langle 0 | \delta (\xi - i a) | \psi    \rangle  &, \quad  |\psi \rangle  & :=  \psi (a^{\dag}) | 0 \rangle
	\end{alignedat}
	\label{fouriertr}
\end{equation}

Appendix \ref{fourier} provides a proof of these relations using the properties of coherent states. This transformation reveals the following correspondence:\footnote{{\cite{Bengtsson:2013vra}} also uses creation and annihilation operators to write the Wigner equations, but the correspondence there is $\xi_{\mu} + i \partial_{\xi} \sim a_{\mu}$, $\xi_{\mu} - i \partial_{\xi} \sim a_{\mu}^{\dag}$ which is in different formulation from ours.
}
\begin{alignat}{3}
      \Psi (\xi) & \leftrightarrow  | \psi \rangle
    & ,\quad  \xi_{\mu} \Psi (\xi) & \leftrightarrow  i a_{\mu} |
    \psi \rangle & ,\quad  \partial_{\xi} \Psi (\xi) & \leftrightarrow  i
    a^{\dag}_{\mu} | \psi \rangle
\end{alignat}
For our convenience, we further perform a Fourier transformation for $\Psi (p,\xi)$ in terms of variables $p^{\mu}$ to spacetime coordinates $x^{\mu}$. Then, we find the Fourier(-like)-transformed Wigner equations written in terms of creation and annihilation operators:
\begin{align}
  \partial^2 | \psi (x) \rangle & =  0   \label{kg_oc}\\
  (a^{\dag} \cdot \partial + i \mu) | \psi (x) \rangle & =  0 
  \label{gauge_oc}\\
  (a \cdot \partial) | \psi (x) \rangle & =  0   \label{div_oc}\\
  (a^2 + 1) | \psi (x) \rangle & =  0 \quad .  \label{trace_oc}
\end{align}
Here, $\partial_{\mu} = \frac{\partial}{\partial x^{\mu}}$, $a \cdot \partial=a^\mu \partial_\mu$, and $a^2=a^\mu a_\mu$. These equations can also be derived by replacing the variables used in {\cite{Bekaert:2005in}} with the following correspondences: $w^{\mu}  \rightarrow  a^{\dag \mu}  ,  \frac{\partial}{\partial w_{\mu}}  \rightarrow  a^{\mu} ,    \Psi  \rightarrow  | \psi (x) \rangle$, along with a Fourier transformation to coordinate space.\footnote{See eq.(2.8) - eq.(2.11) of {\cite{Bekaert:2005in}}.
}

We introduce so-called HS operators($l_1 := a^{\mu} \partial_{\mu}, l_2 := \frac{1}{2} a^{\mu} a_{\mu}$) that are used throughout this paper. See Appendix \ref{convention} for all definitions and properties. Then, eqs.(\ref{kg_oc})-(\ref{trace_oc}) can be written as follows:\footnote{We omit the spacetime coordinate $x$ from now on.
}
\begin{align}
  \partial^2 | \psi \rangle & =  0   \label{kg_hs}\\
  (l_1^{\dag} - i \mu) | \psi \rangle & =  0  \label{gauge_hs}\\
  l_1 | \psi \rangle & =  0   \label{div_hs}\\
  (l_2 + \omega) | \psi \rangle & =  0, \quad \omega = \frac{1}{2} \quad . 
  \label{trace_hs}
\end{align}
Incidentally, both eq.(\ref{kg_hs}), the massless Klein-Gordon equation, and eq.(\ref{div_hs}), the divergence free condition (Lorentz gauge for spin $1$ or de Donder gauge for spin $2$), are shared with HS (helicity) fields. On the other hand, for HS fields, eq.(\ref{trace_hs}) corresponds traceless condition: $l_2 | \psi \rangle = 0$ and condition like eq.(\ref{gauge_hs}) is not necessary as the irreducibility condition.

In addition to these Wigner equations, all Lagrangians, gauge fixing conditions, and all equations appearing in the following sections of this paper are formulated uniformly using HS operators (or creation and annihilation operators) and Fock space. This allows us to make mathematical treatment more rigorous. Furthermore, the use of HS operators with closed algebraic relations clarifies the mathematical structure.

\subsection{Modification of Wigner Equations}

\subsubsection{Trace Generating Operator $Z_X^{\omega}$}

As noted {\cite{Najafizadeh:2017tin}}, Wigner equations can be modified to include a traceless condition for the field. Here, we perform a similar modification for the Fock state vector and the HS operators acting on it. We introduce a trace-fixed state $| \psi^{\omega} \rangle$ with any real number $\omega$ satisfying the following condition:\footnote{$ |\psi^{1 / 2} \rangle \equiv |\psi \rangle$
}
\begin{align}
      (l_2 + \omega) | \psi^{\omega} \rangle & =  0 \quad.
\end{align}
We aim to establish a relation between the state $|\psi^{\omega} \rangle$ for $\omega = \frac{1}{2}$ and the traceless state $|\psi^0 \rangle$ for $\omega = 0$. This relation is expressed in terms of ``trace generating operator'' $Z_X^{\omega}$ as follows:
\begin{alignat}{3}
    | \psi^{\omega} \rangle & =  Z_X^{\omega} | \psi^0 \rangle & , \quad
      Z_X^{\omega} & =  \sum_{m = 0}^{\infty} \frac{(- \omega  l_2^{\dag})^m}{m!} (X)^{\underline{- m}}& , \quad  
      X & :=  N + D / 2
  \label{traceop_def}
\end{alignat}

Here, $N : = a^{\dag \mu} a_{\mu}$ is number operator and $X$ is ``shifted number operator''. Appendix \ref{traceop} summarizes a derivation method, detailed formulas, and useful properties of $Z_X^{\omega} $. Here, we have used the notation $(x)^{\underline{- n}}$, which represents a product of $n$ factors. Refer to Appendix \ref{factorial} for the definition. The central property of $Z_X^{\omega}$ is\footnote{One can determine $Z_X^{\omega}$ by expressing it as an infinite series expansion in terms of $l_2^{\dagger}$ and solving the resulting difference equation for the expansion coefficients to satisfy the given equation (\ref{traceop_formula}).
}
\begin{align}
  (l_2 + \omega) Z_X^{\omega} & =  Z_{X+2}^{\omega} l_2 \quad. 
  \label{traceop_formula}
\end{align}
From this, for a given traceless state$| \psi^0 \rangle$, where $l_2 | \psi^0 \rangle = 0$, it is obvious that
\begin{align}
    (l_2 + \omega) Z_X^{\omega} | \psi^0 \rangle & =  0\quad.
\end{align}
We observe that the factor $Z_X^{\omega}$ is analogous to the ``coherent state generating operator'' $z^{\omega} : = e^{\omega b^{\dagger} / \hslash}$ (see table \ref{table1}). When $Z_X^{\omega}$ operates on a traceless state $| \psi^0 \rangle$, it becomes a trace-fixed state $| \psi^{\omega} \rangle$ with $\omega$. On the other hand, when $z^{\omega}$ operates on the vacuum $| 0 \rangle$, it becomes a coherent state $| \omega \rangle$ with $\omega$. These relationships are parallel to each other. The key difference lies in the algebraic properties: while the commutator of creation and annihilation operators($b, b^{\dag}$) yields a commutative quantity ($\hslash$), the commutator of the HS operators $l_2$ and $l_2^{\dagger}$ yields a non-commutative quantity: shifted number operator $X + 1$. This relationship can be summarized in the following table:

\begin{table}[htbp]
	\centering
	\begin{tabular}{|l|c|l|}
		\cline{1-1}\cline{3-3}
		\begin{tabular}{l}
			coherent state\\
			generating operator
		\end{tabular} $z^{\omega} = e^{\omega b^{\dagger} / \hslash}$  & &trace  generating operator
		$Z_X^{\omega}$(\ref{traceop_def})\\
		\cline{1-1}\cline{3-3}
		$| \omega \rangle = z^{\omega} | 0 \rangle$ &  & $|
		\psi^{\omega} \rangle = Z_X^{\omega} | \psi^0 \rangle$\\
		\cline{1-1}\cline{3-3}
		vacuum: $| 0 \rangle , b | 0 \rangle = 0$ & $\Leftrightarrow$ & traceless state: $|
		\psi^0 \rangle, l_2 | \psi^0 \rangle = 0 $ \\ 
		\cline{1-1}\cline{3-3}	
		coherent state: $| \omega \rangle, (b - \omega) | \omega \rangle = 0$ && trace-fixed state: $| \psi^{\omega} \rangle, (l_2 + \omega) |
		\psi^{\omega} \rangle = 0$ \\ 
		\cline{1-1}\cline{3-3}
		algebra: $[b, b^{\dag}] = \hbar, b \hbar = \hbar b$  &&  algebra:
		$[l_2, l_2^{\dag}] = X + 1, l_2 X = (X + 2) l_2$ \\
		\cline{1-1}\cline{3-3}
	\end{tabular}
	\caption{\label{table1}Analogous correspondence between coherent state and trace-fixed state}
\end{table}

In Appendix \ref{traceop}, we also introduce the $k$-th trace generating operator $Z_{X_{1 - k}}^{\omega}$. Since $Z_{X_{1 - k}}^{\omega}$ is invertible, it allows for redefinition of fields in equations and Lagrangian in a way that modifies traces of arbitrary order. Both the ($k$-th) trace generating operator $Z$ and the coherent state generating operator $z$ are written in exponential (-like) form. We believe that the similarity between them, which we have pointed out here, is interesting both mathematically and physically.

\subsubsection{Modified Wigner equation written by HS operators}

In order to find modified Wigner equation written by creation-annihilation operator, we simply input eq.(\ref{traceop_def}) to Wigner eqs.(\ref{kg_hs})-(\ref{trace_hs}), move $Z_X^{\omega}$ to most left by using commutation relation between HS operators and $Z$s and factor out $Z$s by using the fact that existence of inverse of $Z$s. See Appendix \ref{m_wigner} for detailed calculations. After $N$-dependent phase rotation $| \psi^0 \rangle  =  e^{- i \pi N / 2} | \Phi \rangle$ for avoiding imaginary unit appearing, the result for Modified Wigner equations become as follows.
\begin{align}
  \partial^2 | \Phi \rangle & =  0  \label{kg_mod}\\
  \left( l_1^{\dag} - \mu + \mu l_2^{\dag} \frac{\omega}{X X_1} \right) | \Phi
  \rangle & =  0  \label{gauge_mod}\\
  \left( l_1 - \mu \frac{\omega}{X} \right) | \Phi \rangle & =  0 
  \label{div_mod}\\
  l_2 | \Phi \rangle & =  0 \quad .  \label{trace_mod}
\end{align}
Note that these equations has the same form to {\cite{Najafizadeh:2017tin}}.\footnote{Replacement of the variables in eqs. (2.14) to (2.17) from {\cite{Najafizadeh:2017tin}} : $   \omega^{\mu}  \rightarrow  a^{\dag \mu} $, $ \frac{\partial}{\partial  \omega_{\mu}}  \rightarrow  a^{\mu} $, $p_{\mu} \rightarrow  - i  \partial_{\mu} $, $\varphi (p, \omega) \rightarrow  | \psi^0 \rangle$, together with performing a Fourier transformation to coordinate space, reproduces our eqs.(\ref{kg_mod_app0})-(\ref{trace_mod_app0}), which are the equations before the phase rotation.
}
As in the original Wigner equations, some of these equations are not independent. In fact, it's easy to see that eqs.(\ref{kg_mod}) and (\ref{div_mod}) can be derived from eqs.(\ref{gauge_mod}) and (\ref{trace_mod})(see Appendix \ref{alg_wigner}).

One can check value of quadratic Casimir $C_2$. From Appendix \ref{casimir}, it is calculated as $C_2 | \Phi \rangle  =  2 \mu^2 \omega | \Phi \rangle $. By setting $\omega = 1 / 2$, this gives correctly Casimir eigenvalue for CS case:
\begin{align}
  C_2 | \Phi \rangle & =  \mu^2 | \Phi \rangle \quad . 
\end{align}
Now, similarly to{\cite{Bekaert:2005in}} and {\cite{Najafizadeh:2017tin}}, eq.(\ref{gauge_mod}) is solved with any state vector $| \tilde{\phi} \rangle$ multiplied by delta function as
\begin{align}
  | \Phi \rangle & =  \delta \left( l_1^{\dag} - \mu + \mu l_2^{\dag}
  \frac{\omega}{X X_1} \right) | \tilde{\phi} \rangle \quad . 
  \label{gauge_invsol}
\end{align}
Then, coefficient $\tilde{\phi}_{\mu_1 \ldots \mu_n} (x)$ appeared in the expansion of $| \tilde{\phi} \rangle$ in terms of $a^{\dag}$ is identified with helicity $n$ vector index field($\mu = 0$) in massless HS field theory. 'Bra' state vector $\langle \tilde{\phi} |$ is given by its conjugation: $a^{\dag} \rightarrow a$ with complex conjugation of the coefficient function.\footnote{In fact, one may also use real functions in the resulting Lagrangian.
}
\begin{align}
  | \tilde{\phi} \rangle = \sum_{n = 0}^{\infty} a^{\dag \mu_1} \cdots a^{\dag
  \mu_n} | 0 \rangle \tilde{\phi}_{\mu_1 \ldots \mu_n} (x) & , \quad \langle
  \tilde{\phi} | = \sum_{n = 0}^{\infty} \tilde{\phi}^{\ast}_{\mu_1 \ldots
  \mu_n} (x) \langle 0| a^{\mu_1} \cdots a^{\mu_n} \quad . 
\end{align}
In Appendix \ref{alg_wigner}, once field $| \tilde{\phi} \rangle$ satisfying:
\begin{align}
  \partial^2 | \tilde{\phi} \rangle & =  0  \label{kg_pol}\\
  \left( l_1 - \mu \frac{\omega}{X} \right) | \tilde{\phi} \rangle & =  0 
  \label{div_pol}\\
  l_2 | \tilde{\phi} \rangle & =  0  \label{trace_pol}
\end{align}
is found, then field $| \Phi \rangle$ defined by eq.(\ref{gauge_invsol}) is proved to satisfy all eqs.(\ref{kg_mod})-(\ref{trace_mod}).

In the following section \ref{lag_const}, we will introduce a Lagrangian and demonstrate that its EOM, along with gauge-fixing conditions, lead to eqs.(\ref{kg_pol})-(\ref{trace_pol}). Furthermore, we will show that under gauge transformations that preserve Lagrangian introduced, the field $| \Phi \rangle$ defined in eq.(\ref{gauge_invsol}) remains invariant.\footnote{It is obvious since the same factor in eq.(\ref{gauge_mod}) or eq.(\ref{gauge_invsol}) $\left( l_1^{\dag} - \mu + \mu l_2^{\dag} \frac{\omega}{X X_1} \right)$ will appear in gauge transformation as $\delta | \tilde{\phi} \rangle = \left( l_1^{\dag} - \mu + \mu l_2^{\dag} \frac{\omega}{X X_1} \right) | \tilde{\varepsilon} \rangle $(see eq.(\ref{gauge_tr})).
}
These results will indicate that the Lagrangian describes the dynamics of CS fields in the irreducible representation of the Poincare algebra. 

\section{Fronsdal-like Constrained Lagrangian Formulation for CS theory}\label{lag_const}

This section aims to obtain a Lagrangian that reproduces the modified Wigner equations derived in Section \ref{wigner} using HS operators. We begin with a CS Lagrangian written in terms of spinor index tensor fields with integer spin, constructed by BRST method in four-dimensional flat spacetime {\cite{Buchbinder:2018yoo}}. We then transform this Lagrangian into Fronsdal-like form described by double-traceless vector index fields, including $\mu$. This Fronsdal-like Lagrangian is then generalized to arbitrary dimensions $D$ by demanding $D$-dimensional gauge invariance. Resulted Lagrangian will be found as $\mathcal{L}_1$(\ref{Lag_1f_sec3}). It is written with creation-annihilation operator and its dynamical variable is constrained as double-traceless. We will demonstrate that its gauge transformation, EOM, and gauge fixing conditions (equivalent to the de Donder-type gauge with tracelessness) match the modified Wigner equations derived in Section \ref{wigner}. This Lagrangian exhibits exact consistency with the one proposed by Metsaev {\cite{Metsaev:2016lhs}}. Furthermore, setting $\mu$ to zero leads to perfect agreement with the $D$-dimensional Fronsdal Lagrangian. This Lagrangian $\mathcal{L}_1$ serves as the starting point for constructing an unconstrained form of Lagrangian, which will be derived in Section \ref{lag_unconst}.

\subsection{Transforming Spin Index Tensor Field Lagrangian into Fronsdal-like Lagrangian in Arbitrary Dimension}

Lagrangian for CS theory in flat four-dimensions was constructed using the BRST method {\cite{Buchbinder:2018yoo}}. It involves three totally symmetric fields, which are not subject to any special constraints imposed manually. However, they are automatically traceless due to the construction using spinor index tensor fields. One of the fields can be eliminated using its EOM, resulting in the Lagrangian being expressed as:\footnote{One can clearly see the Hermiticity of Lagrangian $\mathcal{L}_s$ through this matrix expression, which is used throughout this paper.
}
\begin{alignat}{1}
         \mathcal{L}_s   & =  \begin{pmatrix}
      ^s \langle \varphi |   & ^s \langle \varphi_2|
    \end{pmatrix}
      \begin{pmatrix*}[r]
      \partial^2  + L_1^{s \dag} \frac{1}{2 (N^s + 1)} L^s_1 &\quad - L_1^{s \dag} \frac{1}{2 (N^s + 1)} L_1^{s \dag}\\
      - L^s_1 \frac{1}{2 (N^s + 1)} L^s_1 &\quad - \partial^2  + L^s_1 \frac{1}{2 (N^s + 1)} L_1^{s \dag}
   \end{pmatrix*}
  \begin{pmatrix*}[l]
      | \varphi \rangle^s\\
      | \varphi_2 \rangle^s
    \end{pmatrix*}  
\label{Lag_spin}\\
    & \delta  \begin{pmatrix*}[l]
        | \varphi \rangle^s\\
        | \varphi_2 \rangle^s
       \end{pmatrix*}  =   \begin{pmatrix*}[l]
        L_1^{s \dag}\\
        L^s_1
        \end{pmatrix*} | \varepsilon \rangle^s
    \quad.
  \end{alignat}

Here, we have introduced new notation for HS operators (divergence operators $L_1^s$ and $L_1^{s \dagger}$and the number operator $N^s$ as half the sum of number operators from the bar and un-bar creation-annihilation operators):
\begin{equation}
  \begin{alignedat}{5}
    L^s_1 & :=  l^s_1 - \mu &&,&   \quad    l^s_1 N^s & =   (N^s + 1) l^s_1 &&, & \quad [ {l^s_1}^{\dag}, l^s_1 ] & =  [L_1^{s \dag}, L^s_1]  \\
    L_1^{s \dag} & :=  l_1^{s \dag} - \mu &&,&  \quad  N^s l_1^{s \dag} & =  l_1^{s \dag} (N^s +1)  &&,& \quad   &=   2 (N^s + 1) \partial^2  \quad  .
  \end{alignedat} 
\end{equation}
Here, dagger $\dag$(written as $+$ in {\cite{Buchbinder:2018yoo}}) describe Hermitian conjugation. \ $| \varphi \rangle^s$, $| \varphi_2 \rangle^s$, and $| \varepsilon \rangle^s$ are summed over all ranks of spinor index fields with integer spin $\varphi(x) ^{\dot{b} (n)}_{a (n)} $ contracted with un-bar and bar creation operators $c^a$, $\bar{c}_{\dot{a}}$ with spinor indices $a, \dot{a}$. 'Bra' state vectors $^s \langle \varphi | $ and $^s \langle \varphi_2 |  $ are the Hermitian conjugates of $| \varphi \rangle^s$ and $ | \varphi_2 \rangle^s$, respectively. They are as follows:
  \begin{align}
      | \varphi \rangle^s & =  \sum_{n = 0}^{\infty}   c^{a (n)} \bar{c}_{\dot{b} (n)} |0 \rangle^s \frac{1}{n!} {\varphi (x)}^{\dot{b} (n)}_{a (n)}\\
      {}^s \langle \varphi  |   & =  \sum_{n = 0}^{\infty} \frac{1}{n!} 
      {\varphi (x)}^{\ast b (n)}_{\dot{a} (n)}
    {}^s\langle 0| a_{b (n)} {\bar{a}}^{\dot{a} (n)} \quad,
  \end{align} 
and similar form for $\varphi_2, \varepsilon$. Unlike the original paper {\cite{Buchbinder:2018yoo}}, we use the superscript $s$ ('s'pinor index) for Lagrangian, state vectors and HS operators to distinguish them from those in vector indices. See Appendix \ref{convention} and {\cite{Buchbinder:2018yoo}} for the explicit form of $l$s$, N, \bar{N}$, and other conventions.

\subsubsection{Unitary-like Operator bridging Spinor Fock Space to Vector Fock Space and Lagrangian Transformation}

Here, we introduce a unitary-like operator $U$ for transforming state vectors in spinor Fock space (constructed by spinor index creation operators $c^a$, $\bar{c}_{\dot{a}}$), to state vectors in vector Fock space (constructed by vector index creation operators $a^{\dag \mu}$). Making use of this operator, we rewrite the Lagrangian $\mathcal{L}_s$(\ref{Lag_spin}) described by spinor index totally symmetric tensor fields, into a Lagrangian described by vector index totally symmetric tensor fields. Notably, these transformed vector index fields always become traceless. There is a relation between spinor index tensor field $\varphi (x)^{\dot{b}(n)}_{a(n)}$ and vector index field $\varphi (x)_{\mu (n)}$ for arbitrary-rank totally symmetric tensors {\cite{Buchbinder:1998qv}}. Similarly, but with different normalization, we define a unitary-like operator $U$ that relates the spinor Fock space and the traceless vector Fock space as follows:\footnote{Although not necessary for the following sections, this provides an example of how the formulation can be written using explicit components, which may be useful for some readers. \ For a state vector expanded in components with a fixed rank $n$ tensor, with $u_n = 1 / \sqrt{n!^3 2^n}$, the following relations hold between components that transform into each other under $U$ and $U^{\dag}$:  
	\begin{gather}
    | \varphi \rangle_n  =  {a^{\dag}}^{\mu (n)} | 0 \rangle \varphi_{\mu (n)} \frac{1}{\sqrt{n!}}   \qquad \qquad\qquad ,\qquad \quad | \varphi \rangle_n^s  =  c^{a (n)} \bar{c}_{\dot{b} (n)} |0 \rangle^s \frac{1}{n!}  \varphi {(x)^{\dot{b} (n)}}_{a (n)}
\\
    _n \langle \varphi  |    = \frac{1}{\sqrt{n!}} \varphi^{\ast}_{\mu (n)} \langle 0| a^{\mu (n)}  \quad\qquad\qquad\qquad  ,\qquad \quad  ^s_n \langle \varphi  |    =  \frac{1}{n!}   \varphi {{(x)^{\ast b (n)}}_{\dot{a} (n)}}^s \langle 0| a_{b (n)} {\bar{a}^{\dot{a} (n)}} 
\\
      \varphi_{\mu (n) }  = \frac{(-1)^n}{\sqrt{2}^n} (\sigma_{\mu_1} )_{a_1\dot{b}_1} \ldots (\sigma_{\mu_n} )_{a_n \dot{b}_n} \varphi ^{a (n) \dot{b} (n)}   ,\quad \varphi^{a (n)  \dot{b} (n)}  =  \frac{1}{\sqrt{2}^n} \left( {\bar{\sigma}^{\mu_1}}  \right)^{\dot{b}_1 a_1} \ldots \left( {\bar{\sigma}^{\mu_n}}  \right)^{\dot{b}_n a_n} \varphi_{\mu (n) }
\\
    {}_n \langle \varphi ' | \varphi \rangle_n  =  {\varphi'}^{\ast \mu (n)} \varphi_{\mu (n)}   \quad =\quad  {\varphi'}^{\ast a (n)}_{\dot{b} (n)} \varphi_{a (n)}^{\dot{b} (n)}  =  {}^s_n \langle \varphi'  | \varphi \rangle_n^s \quad . 
    \end{gather}
In the convention of {\cite{Buchbinder:1998qv}}, the following replacements are made with a different normalization: $u_n \rightarrow 1 / (2^n n!^{3 / 2})$, $u_n^{\ast} \rightarrow 1 / (n!^{3 / 2})$, $\varphi_{\mu (n) } \rightarrow \sqrt{2}^n \varphi_{\mu (n) }$. 
}
\begin{equation}
  \begin{aligned}
    U & =  \sum_n u_n a^{\dag \mu (n)} | 0 \rangle (\sigma_{\mu_1} )^{a_1}_{\dot{b}_1}   
    \ldots (\sigma_{\mu_n} )^{a_n}_{\dot{b}_n} {}^s\langle  0|
    a_{a (n)}  \bar{a}^{\dot{b} (n)}\\
    U^{\dag} & =  \sum_n u_n^{\ast} c^{a (n)} \bar{c}_{\dot{b} (n)} |0
    \rangle^s \left( {\bar{\sigma}^{\mu_1}}  \right)^{\dot{b}_1}  _{a_1}
    \ldots \left( {\bar{\sigma}^{\mu_n}}  \right)^{\dot{b}_n}  _{a_n} \langle 
    0| a _{\mu (n)}\quad.
  \end{aligned} 
\label{unitaryop}
\end{equation}
$U$ maps any ``spinor Fock state'' ($\equiv$state vector in spinor Fock space) $| \varphi \rangle^s$ to a ``traceless vector Fock state'' ($\equiv$traceless state vector in vector Fock space) $| \varphi \rangle$, while $U^{\dag}$ maps any traceless vector Fock state $| \varphi \rangle$ to a spinor Fock state $| \varphi \rangle^s$. The normalization factor $u_n$ is determined by the condition that it makes spinor Fock state $| \varphi \rangle^s$ invariant under $U^{\dag} U$:
\begin{align}
    U^{\dag} U | \varphi \rangle^s &=  | \varphi \rangle^s \quad.
\end{align}
This condition determines absolute value $|u_n | = 1 / \sqrt{n!^3 2^n}$. Under this condition, traceless vector Fock state $| \varphi \rangle$ is invariant under $U^{\dag} U$:
  \begin{align}
      U U^{\dag} | \varphi \rangle &=  | \varphi \rangle \quad.
  \end{align} 
We denote state vectors that interchange under $U$ and $U^{\dag}$ by the same name, such as \ $| \varphi \rangle$ and $| \varphi \rangle^s$. Determining $|u_n |$ as noted above ensures the following relationships hold:
  \begin{align}
    U | \varphi \rangle^s  = & | \varphi \rangle\\
    U^{\dag} | \varphi \rangle  = & | \varphi \rangle^s \quad.
  \end{align}
Furthermore, by choosing $u_n$ to be real (e.g., $u_n = 1 / \sqrt{n!^3 2^n}$), both the components of $| \varphi \rangle$ and $| \varphi \rangle^s$ can be chosen as real as well.

When rewriting the Lagrangian $\mathcal{L}_s$(\ref{Lag_spin}), we can insert $U^{\dagger} U$ not only before spinor Fock state $| \varphi \rangle^s$ but also before operator, as if it were an identity operator. This can be justified as follows. Suppose we temporarily define $O$s as functions of $l_1^s$, $l_1^{s \dagger}$. Since $O| \varphi \rangle^s$ is in the same Fock space as $| \varphi \rangle^s$, it can be written as $|O \phi \rangle^s$. Then, we have:
\begin{align}
  U^{\dagger} UO| \varphi \rangle^s &= U^{\dagger} U|O \varphi \rangle^s = |O  \varphi \rangle^s = O| \varphi \rangle^s \quad .
\end{align}
Therefore, we can insert \ $U^{\dagger} U$ before $O | \varphi \rangle^s$. By writing $f (N^s)$ as any function of $N^s$, terms in the Lagrangian $\mathcal{L}_s$(\ref{Lag_spin}) are rewritten as follows:
  \begin{align}
    ^s \langle \varphi | O_1 f (N^s) O_2 | \varphi \rangle^s   & =  \langle \varphi | (U O_1 U^{\dag}) f (N) (U O_2 U^{\dag}) | \varphi \rangle  \quad .
  \end{align} 
Here, property $  U N^s U^{\dag}  =  N U U^{\dag}$ is used. When calculating directly, we find that the expression in the form of $U O_2 U^{\dag} | \varphi \rangle$ can be written as follows:
  \begin{align}
      U l_1^s U^{\dag} | \varphi \rangle & = - \sqrt{2 (N + 1)} \left( l_1 + \frac{1}{N + 1} l_1^{\dag} l_2 \right) | \varphi \rangle
  \\
      U l_1^{s \dag} U^{\dag} | \varphi \rangle & =  - \left( l_1^{\dag} + l_2^{\dag} l_1 \frac{1}{N + 1} \right) \sqrt{2 (N + 1)} | \varphi       \rangle \quad.
  \end{align} 
Here, $l_{1, 2}^{(\dag)}$ are HS operators in vector index(see Appendix \ref{convention}).

Incidentally, through an examination of these coefficients, we introduce $l_1^{\prime}$  (and its conjugate $l_1^{\prime \dagger}$) for arbitrary spacetime dimensions $D$ as:
  \begin{align}
    l_1' & =  - \sqrt{2 X} \left( l_1 + \frac{1}{X} l_1^{\dag} l_2 \right)\quad .
  \end{align} 
Here, the shifted number operator $ X =  N + D / 2 - 1$ was introduced in Section \ref{wigner}. It can be verified that the relative coefficient of the terms in $l_1'$ is uniquely determined by requiring that the commutation relation $[l_1' {, l_1'}^{\dag}]$ on the traceless Fock space has the same form as $[l_1^{s \dag}, l^s_1]$.\footnote{In fact, it is calculated as follows, using $X = N + D / 2 - 1$:
	\begin{align}
    {}[l_1' {, l_1'}^{\dag}] & = 2 X \partial^2 - \frac{2}{X} l_2^{\dag} l_2 \partial^2 + \frac{2}{X_{- 1} X} l_2^{\dag} l_1^{\dag} l_1 l_2 \quad.
  \end{align} 
From the tracelessness of $| \varphi \rangle$, one finds $[l_1' ,l_1'^\dag] | \varphi \rangle = 2 X \partial^2 | \varphi \rangle \xrightarrow[D \rightarrow 4]{} 2 (N + 1) \partial^2 | \varphi \rangle$.
}

The newly introduced unitary-like operator $U$ allows for a systematic transformation between spinor Fock space and vector Fock space. We believe  this can be useful not only for the present study but also for a wide range of studies involving higher-rank tensors.

Following the above rewriting procedure and straightforward calculation, Lagrangian written in vector Fock space for four-dimensional case is given as follows:
  \begin{gather}
  	\mathcal{L}_2	= \langle \overrightarrow{\phi_2} |  
  	\begin{pmatrix*}[c] \partial^2  + L_X^{\dag} L_X & - L_X^{\dag} \frac{1}{\sqrt{X}} L_X^{\dag} \sqrt{X}
    \\
  		- \sqrt{X} L_X \frac{1}{\sqrt{X}} L_X & - \left( 1 + \frac{X}{X_1} \right) \Delta + \frac{1}{X_1} L_X^{\dag} X X_2 L_X   		\frac{1}{X_1} 
  	\end{pmatrix*}| \overrightarrow{\phi_2} \rangle  \label{Lag_2f}
\qquad\\
  	\delta | \overrightarrow{\phi_2} \rangle   =  
  	\begin{pmatrix*}[l]
  		L_X^{\dag} + l^{\dag}_2 l_1 \frac{1}{X}
     \\
 		\sqrt{X} L_X \frac{1}{\sqrt{X}} 
  	\end{pmatrix*}\sqrt{2 X} | \varepsilon \rangle ,\quad 	l_2 | \overrightarrow{\phi_2} \rangle   =   l_2 | \varepsilon \rangle =  0,\quad
  	| \overrightarrow{\phi_2} \rangle   : =  
  	\begin{pmatrix*}[l]
  		| \varphi \rangle
  	\\
  		| \varphi_2 \rangle 
  	\end{pmatrix*}  
\qquad\\
  	\qquad \quad L_{X_k}  :=  l_1 - \frac{\mu}{\sqrt{2 X_k}}  , \quad \Delta  :=  \partial^2 - \frac{\mu^2}{2 X X_1 (X + X_1)}  ,\quad
  	\begin{matrix*}[l]
  		X_k  = X + k  ,  k \in \mathbb{Z}
  	\\
  		X  =  N + D / 2- 1  ,  D = 4 \quad . 
  	\end{matrix*}
  \end{gather}

This is a $\mu$-extended version of the Fronsdal Lagrangian expressed by two traceless fields. While it can be easily checked that it may be regarded as a $D$-dimensional gauge invariant Lagrangian, we will focus on deriving a Lagrangian with a single double-traceless field to approach the purpose of this paper.

\subsubsection{Fronsdal-like Lagrangian with Single Double-Traceless Field for CS Theory in Arbitrary Spacetime Dimension}

To transform the Lagrangian $\mathcal{L}_2$(\ref{Lag_2f}), which is written in terms of two traceless fields $| \varphi \rangle$and $ | \varphi_2 \rangle$, into a Lagrangian expressed in terms of a single double-traceless field, we can introduce a double-traceless field $ | \phi \rangle$ as follows:
\begin{align}
        | \phi \rangle & :=  | \varphi \rangle - l^{\dag}_2 \frac{1}{\sqrt{X X_1}} | \varphi_2 \rangle \quad .
\end{align}
Conversely, it is possible to express the two traceless fields in terms of the double-traceless field:
\begin{alignat}{2}
	\begin{pmatrix*}[l]
    | \varphi \rangle
    \\
        | \varphi_2 \rangle
	\end{pmatrix*}
      & 
 =  \begin{pmatrix*}[c]
        \Pi_1
        \\
        - \sqrt{X / X_1} l_2
    	\end{pmatrix*}
| \phi \rangle &, \quad  \Pi_1 & :=  1 - l_2^{\dag} \frac{1}{X_1}  l_2 \quad.
\label{2fto1f}
\end{alignat}
Here, the operator $\Pi_1$ is a projection operator onto traceless fields within double-traceless Fock space.\footnote{See also Appendix \ref{convention} for the projection operator $\Pi_1$.
}
From these relations, we can see the following equivalence:
\begin{alignat}{2}
    l_2^2 | \phi \rangle & =  0
	& \quad \Leftrightarrow \quad  l_2 
	\begin{pmatrix*}[l]
      | \varphi \rangle
	\\
      | \varphi_2 \rangle
	\end{pmatrix*} & =  0
 \quad . 
\end{alignat}
Substituting eq.(\ref{2fto1f}) into $\mathcal{L}_2$(\ref{Lag_2f}) results in Lagrangian as follows:
\begin{gather}
          \mathcal{L}_1
        = \langle \phi | 
        \begin{pmatrix}
          1 & l^{\dag}_2
        \end{pmatrix} 
        \begin{pmatrix*}[c]
          \partial^2  + L_X^{\dag} L_X & L_X^{\dag} L_{X_1}^{\dag}
        \\
          L_{X_1} L_X & - 2 \partial^2  + L_{X_2}^{\dag} L_{X_2}
        \end{pmatrix*} 
        \begin{pmatrix}
          1\\
          l_2
        \end{pmatrix} | \phi \rangle 
\label{Lag_1f_sec3}\\
        \delta | \phi \rangle = \tilde{L}_X^{\dag} \sqrt{2 X} |\varepsilon \rangle , \quad l_2^2 | \phi \rangle  = l_2 | \varepsilon \rangle = 0,\quad 
      \tilde{L}_X  : =  L_X + \frac{\mu}{X \sqrt{2 X_1}} l_2\quad. \label{gauge_1f_sec3}
\end{gather}

The Lagrangian $\mathcal{L}_1$(\ref{Lag_1f_sec3}) has only been examined in four-dimensions up to now. However, there is an expectation for its extension to arbitrary $D$-dimensions. We present $\mathcal{L}_1$(\ref{Lag_1f_sec3}) as Lagrangian for CS theory in arbitrary $D$-dimensions by treating $D$ as arbitrary in $X$:
\begin{align}
      X & =   N + D / 2 - 1  , \quad D : \text{arbitrary spacetime dimension}
\end{align}
To confirm its validity, it is necessary to first ensure that the Lagrangian $\mathcal{L}_1$ remains invariant under gauge transformations in $D$-dimensions, which can be directly verified through calculation. Moreover, considering correspondences with other literature, it can be observed that the Lagrangian $\mathcal{L}_1$ coincides with the Lagrangian in Metsaev {\cite{Metsaev:2016lhs,Metsaev:2018lth}} by the relation:
\begin{align}
      | \phi_{\text{Metsaev}}\rangle & = \frac{v^N}{\sqrt{N!}} | \phi \rangle  ,\quad  \kappa  =  \mu
 \end{align}
Here, $ | \phi_{\text{Metsaev}}\rangle$ , $v$ and $\kappa$ represent Metsaev's field, additional oscillator, and CS parameter, respectively.

On the other hand, when $\mu$ is set to zero, the Lagrangian $\mathcal{L}_1$ exactly coincides with the $D$-dimensional Fronsdal Lagrangian {\cite{Fronsdal:1978rb}}. In fact, the Lagrangian $\mathcal{L}_1$ can be expanded in terms of $\mu$ as follows:
\begin{align}
    \mathcal{L}_1 & =  \langle \phi | 
    \begin{pmatrix}
	1 & l^{\dag}_2
\end{pmatrix} 
\begin{pmatrix*}[c]
\partial^2_{} + l_1^{\dag} l_1, & {l^{\dag}_1}^2\\
    l_1^2, & - 2 \partial^2_{} + l_1^{\dag} l_1
\end{pmatrix*} 
\begin{pmatrix}
	1\\
	l_2
\end{pmatrix} | \phi \rangle 
   + \mu \text{ dependent terms} .
\end{align}
The first term, which is independent of $\mu$, exactly reproduces the Fronsdal Lagrangian. This shows that (\ref{Lag_1f_sec3}) is a natural extension of the Fronsdal Lagrangian, providing suggestive insights into the extension of operators:
\begin{align}
  l_1 & \xrightarrow[\text{extend}]{} L_{X_k} \quad . 
\end{align}
Here $L_{X_k}$, which we call a generalized divergence operator, reduces to $l_1$ as $\mu \rightarrow 0$:
\begin{align}
  L_{X_k} & \xrightarrow[\mu \rightarrow 0]{} l_1 = a^{\mu} \partial_{\mu}
  \quad . 
\end{align}
In all subsequent Lagrangians, the HS operator $l_1$ used when $\mu = 0$ is always replaced by $L$s including $\mu$ implicitly.\footnote{See eq.(\ref{div_op}) in Appendix \ref{convention} for other generalized divergence operators, all of which reduce to $l_1$ as $\mu \rightarrow 0$.
}
Therefore, the overall forms of the Lagrangians are similar to those with $\mu = 0$. This formulation of our Lagrangians clearly demonstrates that it is a natural extension of the corresponding Lagrangians with $\mu = 0$.

\subsection{Deriving Modified Wigner Equations via Gauge Fixing Equivalent to de Donder-Type}

The crucial point, for Lagrangian to correctly represent CS theory, is to confirm that Wigner equations in $D$-dimensions emerge from the Lagrangian. Here, we explain how Wigner equations emerge by considering the form of EOM and gauge transformations derived from the Lagrangian $\mathcal{L}_1$(\ref{Lag_1f_sec3}), along with the appropriate choice of gauge fixing for the fields.

\subsubsection{Gauge Fixing Procedure}

By examining the form of EOM(see Appendix \ref{eom_frions}) derived from the Lagrangian $\mathcal{L}_1$(\ref{Lag_1f_sec3}), it becomes evident that under the appropriate choice of the following two gauge fixing conditions:
\begin{align}
      L_X | \phi \rangle & =  0 \label{gauge_fixing1}\\
      l_2 | \phi \rangle & =  0 \label{gauge_fixing2} \quad,
\end{align}
the EOM simplify to $\partial^2 | \phi \rangle = 0$.

Indeed, in order to implement this gauge fixing technique, by looking at the form of gauge transformation of eqs.(\ref{gauge_1f_sec3}), one can choose a traceless gauge parameter $ | \varepsilon \rangle$ from the solutions of the following equations:
\begin{align}
      \sqrt{2 X}\partial^2 | \varepsilon \rangle & =  L_X | \phi \rangle\\
      \sqrt{2 X_1}  L_X | \varepsilon \rangle
      & =  l_2  | \phi \rangle
\end{align}
Interestingly, even with the presence of the CS parameter $\mu$, after this gauge fixing, the fields $| \phi \rangle$ and gauge parameter $|\varepsilon \rangle$ follow exactly the same equations:
\begin{alignat}{3}
  \partial^2 | \phi \rangle &= \partial^2 | \varepsilon \rangle && =  0\\
  L_X | \phi \rangle &= L_X | \varepsilon \rangle && =  0  \\
  l_2  | \phi \rangle &= l_2 | \varepsilon \rangle && =  0 \quad . 
\end{alignat}
The relation to the de Donder-type gauge is seen as follows. The Lagrangian $\mathcal{L}_1$(\ref{Lag_1f_sec3}) can be rewritten as\footnote{This could be expressed more similarly to {\cite{Metsaev:2016lhs,Najafizadeh:2017tin}} as:
\begin{align}
  L_{\text{(de Donder)}}
 & =  l_1 + l_1^{\dag} l_2 + \mu \left( \frac{1}{\sqrt{2 X}} \Pi_1 + \frac{1}{\sqrt{2 X_1}} l_2 \right) \quad . 
\end{align}
}(see eq.(\ref{EOMop_formula}) in Appendix \ref{convention}):
\begin{align}
       & \mathcal{L}_1  =  \langle \phi | \left\{ (1 - l^{\dag}_2 l_2) \partial^2 + L_{\text{(de Donder)}}^\dag L_{\text{(de Donder)}} \right\}  | \phi \rangle
\\
      & L_{\text{(de Donder)}}  : =  L_X + \tilde{L}_X^{\dag} l_2 \quad.
      \label{Lag_1_deDonder}
\end{align}
The de Donder-type gauge condition is now written as:
\begin{align}
    L_{\text{(de Donder)}}| \phi \rangle & =  0 \quad .
\label{gauge_fixing_dtrace}
\end{align}
This is proposed by Metsaev in {\cite{Metsaev:2016lhs,Metsaev:2008fs}}. Together with the traceless condition $ l_2 | \phi \rangle = 0$, it forms a gauge fixing in {\cite{Metsaev:2016lhs,Najafizadeh:2017tin}} that is equivalent to ours: (\ref{gauge_fixing1}) and (\ref{gauge_fixing2}).

\subsubsection{Derivation of the Modified Wigner Equations}

In this way, EOM, two gauge fixing conditions, and gauge transformation from the Lagrangian $\mathcal{L}_1$(\ref{Lag_1f_sec3}) can be written as follows:
\begin{align}
        \partial^2 | \phi \rangle & =  0 \label{preWig_kg}\\
        \delta | \phi \rangle & = \tilde{L}_X^{\dag} \sqrt{2 X} |
        \varepsilon \rangle \label{preWig_gauge}\\
        L_X | \phi \rangle & =  0 \label{preWig_div}\\
        l_2 | \phi \rangle & =  0\quad.
\label{preWig_trace}
\end{align}
To demonstrate the equivalence with Wigner equations, let's perform a number operator-dependent rescaling of $ \phi \rangle$ and gauge parameter $| \varepsilon \rangle$ as follows:
\begin{alignat}{2}
      | \phi \rangle & =  \prod_{k = 1, 2, 3, \ldots} \sqrt{2 X_{- k}} |
      \tilde{\phi} \rangle & ,\quad  | \varepsilon \rangle & =  \prod_{k = 1, 2, 3, \ldots} \sqrt{2 X_{- k}} | \tilde{\varepsilon} \rangle \quad.
\end{alignat}
Under this rescaling, eqs.(\ref{preWig_kg})-(\ref{preWig_trace}) transform into:
\begin{align}
  \partial^2 | \tilde{\phi} \rangle & =  0  \label{kg_pol2}\\
  \delta |\tilde{\phi} \rangle & =  \left( l_1^{\dag} - \mu +  l^{\dag}_2 \frac{\mu}{2 X X_1} \right) | \tilde{\varepsilon} \rangle  \label{gauge_tr} \\
  \left( l_1 - \frac{\mu}{2 X} \right) | \tilde{\phi} \rangle & =  0 
  \label{div_pol2}\\
  l_2 | \tilde{\phi} \rangle & =  0 \quad .  \label{trace_pol2}
\end{align}
From eq.(\ref{gauge_tr}), it is evident that the field defined as
\begin{align}
  | \Phi \rangle & =  \delta \left( l_1^{\dag} - \mu + l_2^{\dag}
  \frac{\mu}{2 X X_1} \right) | \tilde{\phi} \rangle \quad ,
  \label{gauge_invsol2}
\end{align}
becomes gauge invariant due to the property of the delta function, $x \delta (x) \equiv 0$. This property of $\Phi$ can also be expressed as an equation:
\begin{align}
  \left( l_1^{\dag} - \mu + l_2^{\dag} \frac{\mu}{2 X X_1} \right) | \Phi
  \rangle & = 0 \quad . 
\end{align}
Thus, by utilizing the definition (\ref{gauge_invsol2}) of $| \Phi \rangle$, when the three eqs.(\ref{kg_pol2}), (\ref{div_pol2}), and (\ref{trace_pol2}) hold for $| \tilde{\phi} \rangle$, as discussed in Section \ref{wigner} and Appendix \ref{alg_wigner}, it becomes evident that the following four equations are satisfied for $| \Phi \rangle$:
\begin{align}
  \partial^2 | \Phi \rangle & =  0 \\
  \left( l_1^{\dag} - \mu + l_2^{\dag} \frac{\mu}{2 X X_1} \right) | \Phi
  \rangle & =  0 \\
  \left( l_1 - \frac{\mu}{2 X} \right) | \Phi \rangle & =  0 \\
  l_2 | \Phi \rangle & =  0 \quad . 
\end{align}
These represent modified Wigner eqs.(\ref{kg_mod})-(\ref{trace_mod}) in $D$-dimensions discussed in Section \ref{wigner}. Thus, the derivation of Wigner equations from the Lagrangian $\mathcal{L}_1$(\ref{Lag_1f_sec3}) implies that the $\mathcal{L}_1$ represents CS theory in $D$-dimensions.
\subsubsection{Comment on Direct Derivation of Original Wigner equation}

At the end of this section, we comment on the possibility of directly deriving the un-modified Wigner equations.

We know $k$-th trace generating operator $Z_{X_{1- k }}^{\omega}$ (see Appendix \ref{traceop}). Using this operator, we can rewrite the double-traceless field and traceless gauge parameter in Lagrangian (\ref{Lag_1_deDonder}) as follows (see eq.(\ref{traceop_prop})):
\begin{alignat}{4}
      (l_2 + \omega)^2 | \phi^{\omega} \rangle & =  0 & ,\quad  | \phi^{\omega}
      \rangle & =  Z_{X_{- 1}}^{\omega} | \phi \rangle & ,\quad  l_2^2 | \phi
      \rangle & =  0\\
      (l_2 + \omega) | \varepsilon^{\omega} \rangle & =  0 & ,\quad  |\varepsilon^{\omega} \rangle & =  Z_X^{\omega} | \varepsilon \rangle & ,\quad  l_2 | \varepsilon \rangle  &=  0,\quad \omega = \frac{1}{2} \quad .
\end{alignat}
Since $Z_{X_k}^{\omega}$ has an inverse (see eq.(\ref{traceop_inv_def})), we can write:
\begin{alignat}{2}
    | \phi \rangle & = (Z_{X_{-1}}^{\omega})^{-1} |\phi^{\omega} \rangle & ,\quad  | \varepsilon \rangle & =  (Z_X^{\omega})^{- 1} |    \varepsilon^{\omega} \rangle \quad.
\end{alignat}

Substituting these into Lagrangian (\ref{Lag_1_deDonder}) and the gauge transformation, we can formally write:
\begin{align}
      \mathcal{L}_1 & =  \langle \phi^{\omega}
      | (Z_{X_{-1}}^{\omega \dag})^{-1}
      \left\{ 
     	 (1-l^{\dag}_2 l_2) \partial^2 +
   		 L_{\text{(de Donder)}}^\dag 
   		 L_{\text{(de Donder)}}
     \right\}
    	(Z_{X_{-1}}^{\omega})^{-1} |\phi^{\omega}\rangle\\    	
    \delta | \phi^{\omega} \rangle & =  Z_{X_{- 1}}^{\omega} \tilde{L}_X^{\dag} \sqrt{2 X} (Z_X^{\omega})^{-1} |\varepsilon^{\omega}\rangle
\end{align}
This is a Lagrangian using a double-trace-full constrained field $| \phi^{\omega} \rangle $ and a trace-full constrained gauge parameter $| \varepsilon^{\omega} \rangle$. It is interesting that the unmodified Wigner eqs.(\ref{kg_hs})-(\ref{trace_hs}) may be derived directly from this Lagrangian.

\section{Unconstrained Lagrangian Formulation for CS theory}\label{lag_unconst}

This section aims to obtain unconstrained form of Lagrangians for CS theory that are equivalent to the Lagrangian $\mathcal{L}_1$(\ref{Lag_1f_sec3}) derived in Section \ref{lag_const}, under partial use of gauge-fixing and EOM. The formulation of unconstrained Lagrangians for HS or CS theory has different types. For instance, in the formulation known as geometric approach by Francia Sagnotti {\cite{Francia:2005bu,Francia:2002aa,Francia:2002pt,Francia:2006hp}}, Lagrangian includes non-local or higher-order derivative terms. On the other hand, methods such as solving the double-trace constraint by Segal {\cite{Segal:2001qq}} and Schuster-Toro {\cite{Schuster:2014hca}} introduce delta function arising from the solution into Lagrangian {\cite{Najafizadeh:2018cpu}}. However, our goal is to derive a more conventional Lagrangian in field theory. The targeted Lagrangian avoids non-local nor delta functions. Instead, we allow auxiliary fields into Lagrangian.

One typical method to obtain this type of Lagrangian is BRST construction {\cite{Buchbinder:1998qv,Burdik:2001hj,Buchbinder:2001bs,Buchbinder:2002ry,Bekaert:2003uc,Buchbinder:2004gp,Sagnotti:2003qa}}. Lagrangians in unconstrained forms are constructed using this method {\cite{Buchbinder:2001bs,Buchbinder:2004gp,Buchbinder:2005ua,Buchbinder:2006nu,Buchbinder:2006ge,Fotopoulos:2008ka}}, typically involving several auxiliary fields. By partially fixing gauge and using EOM, the number of auxiliary fields can be reduced while maintaining unconstrained nature. Lagrangians with the minimum number of auxiliary fields and without higher-order derivatives, while maintaining unconstrainedness, have been obtained for both massive and massless cases with $\mu = 0$ {\cite{Buchbinder:2007ak,Buchbinder:2008ss}}. By allowing for constraints and proceeding with this procedure, the number of fields can be further reduced, resulting in a Fronsdal-like Lagrangian. In other words, within this procedure's context, {\cite{Buchbinder:2007ak,Buchbinder:2008ss}} can be regarded as intermediate between Lagrangians derived from the BRST construction method and those of Fronsdal type. 

In this section, we will firstly obtain a version of {\cite{Buchbinder:2007ak}} extended to include non-zero $\mu$. However, our approach does not start with BRST Lagrangian but with the Fronsdal-like Lagrangian, i.e., starting from the Lagrangian $\mathcal{L}_1$(\ref{Lag_1f_sec3}) and following the reverse of the procedure mentioned above. In other words, starting from $\mathcal{L}_1$, we introduce auxiliary fields to eliminate constraints on fields and gauge parameter. Here, the term ``auxiliary fields'' includes fields that undergo gauge transformations. As a result, a total of five fields, including four auxiliary fields, are required. Secondary, by allowing the third-order derivative terms, we also obtain an unconstrained form of Lagrangian, which is an extension of {\cite{Francia:2005bu}}. It is found to include three fields, the same as in the case of $\mu = 0$. These two derived Lagrangians describe CS fields in the irreducible representation. This is because they can derive Wigner eqs.(\ref{kg_mod})-(\ref{trace_mod}) obtained in Section \ref{wigner} through the Lagrangian $\mathcal{L}_1$(\ref{Lag_1f_sec3}) obtained in Section \ref{lag_const}.

\subsection{Lagrangian with Unconstrained Fields and a Traceless Gauge Parameter}

We will start with the Fronsdal-like Lagrangian $\mathcal{L}_1$(\ref{Lag_1f_sec3}) obtained in Section \ref{lag_const}, using the double-traceless fields:
\begin{gather}
	\mathcal{L}_1 = \langle s_1 | \begin{pmatrix}
		1 & l^{\dag}_2
	\end{pmatrix} 
	\begin{pmatrix*}[c]
		\partial^2  + L_X^{\dag} L_X & L_X^{\dag} L_{X_1}^{\dag}\\
		L_{X_1} L_X & - 2 \partial^2  + L_{X_2}^{\dag} L_{X_2}
	\end{pmatrix*} 
	\begin{pmatrix}
		1\\
		l_2
	\end{pmatrix} | s_1 \rangle 
\label{Lag_1f}\\
\begin{aligned}
	& \delta | s_1 \rangle = \tilde{L}_X^{\dag}  |\lambda \rangle , \quad &l_2^2 | s_1 \rangle  = l_2 | \lambda \rangle = 0 \qquad&.\qquad\qquad\qquad\quad
\end{aligned}
\end{gather}

Here, we have slightly changed notations: $\phi$ $\rightarrow$ $s_1$, $\sqrt{2 X} \epsilon$ $\rightarrow$ $\lambda$.\footnote{The notations $s_i, a_i, \lambda$ used in this section are renamed to correspond to the notations used in the BRST formulated Lagrangian {\cite{Buchbinder:2005ua,Buchbinder:2006nu,Buchbinder:2006ge}}. The fields and gauge parameters appearing there correspond to the symbols in this section's notation when the Lagrangian is reduced to the one with the minimum number of auxiliary fields, as in {\cite{Buchbinder:2007ak,Buchbinder:2008ss}}.
}
In order to eliminate constraints for field $s_1$, we introduce three new unconstrained fields $s_2$, $a_2$, $a_3$. Here the EOM of $a_2$ and $a_3$ should be, respectively, $ |s_2 \rangle =  - l_2 |s_1 \rangle$ and $l_2 |s_2 \rangle = 0$ so that double tracelessness of $s_1$ is derived by their EOM. It means that $s_1$ also can be treated as unconstrained field in Lagrangian. This part of the procedure is similar to that of {\cite{Buchbinder:2007ak,Buchbinder:2008ss}} and the CS parameter $\mu$ does not pose a difficulty in this process. The Lagrangian is given as follows:
\begin{alignat}{1}
	\mathcal{L}_4 &=
		\begin{pmatrix}
		\langle s_1 | & \langle s_2 |
		\end{pmatrix} 
		\begin{pmatrix*}[c]
		\partial^2  + L_X^{\dag} L_X & -L_X^{\dag} L_{X_1}^{\dag}\\
		-L_{X_1} L_X & - 2 \partial^2  + L_{X_2}^{\dag} L_{X_2}
		\end{pmatrix*} 
		\begin{pmatrix}
		| s_1 \rangle \\
		| s_2 \rangle 
		\end{pmatrix} 
	\notag\\
	&+  \left\{
		\begin{pmatrix}
		\langle a_2 |  & \langle a_3 | 
		\end{pmatrix}
		\begin{pmatrix}
		l_2 & 1\\
		0 & - l_2
		\end{pmatrix}
		\begin{pmatrix}
		|s_1 \rangle\\
		|s_2 \rangle
		\end{pmatrix}+ c.c. 
	\right \}\qquad \qquad \qquad
	\label{Lag_4f}\\
&\begin{alignedat}{3}
	\delta |s_1 \rangle  &=  \tilde{L}_X^{\dag} | \lambda \rangle &, 
	\quad \delta |a_2 \rangle  &=  - (\partial^2_{} + L_{X_2}^{\dag} L_{X_2}) 
	\frac{\mu}{X \sqrt{2 X_1}} | \lambda \rangle &&\\
	\delta |s_2 \rangle  &=  \sqrt{X_1} L_X \frac{1}{\sqrt{X}} | \lambda \rangle  &, \quad \delta | a_3 \rangle  &=  - L_{X_3} L_{X_2}  \frac{\mu}{X \sqrt{2 X_1}} | \lambda \rangle &,\quad  l_2| \lambda \rangle &= 0 \quad .	
\end{alignedat}
\end{alignat}

Note that, unlike $\mu = 0$ case, $a_2$, $a_3$ transform under gauge transformation.\footnote{It's worth mentioning that there is a method, suggested by V.A. Krykhtin, where $a_3$ does not transform under gauge transformation. This is achieved by eliminating a certain component proportional to $s_2$ from $a_3$. However, we do not adopt this approach here because, in that case, we could not find a way to construct a final unconstrained form of \ Lagrangian without introducing higher-order derivative terms.
}
One can verify gauge invariance through direct calculation. 

The equivalence between $\mathcal{L}_4$(\ref{Lag_4f}) and $\mathcal{L}_1$(\ref{Lag_1f}) is demonstrated through the equivalence of their EOM. The equivalence of the EOM associated with $\mathcal{L}_4$ and $\mathcal{L}_1$ can be proved as follows (see Appendix \ref{eom_unconstl} for details). To begin, the EOM of $\mathcal{L}_4$ consist of four equations. Using the EOM of $\langle a_3 |$, $|s_2 \rangle$ is tracelessly constrained. Subsequently, using the EOM of $\langle a_2 |$, $|s_2 \rangle$ can be expressed as
\begin{align}
  |s_2 \rangle &= - l_2 |s_1 \rangle,
\end{align}
imposing double-traceless constraint on $|s_1 \rangle$ simultaneously. The remaining two equations reveal that $|a_2 \rangle$ and $|a_3 \rangle$ can also be expressed in terms of $|s_1 \rangle$. The equation concerning $ |s_1 \rangle$ that remains is confirmed to coincide with the EOM of $\mathcal{L}_1$.

\subsection{Lagrangian with Unconstrained Fields and an Unconstrained Gauge Parameter}

In this subsection, we aim to remove constraint on the gauge parameter and derive a fully unconstrained form of Lagrangian. We first derive a conventional Lagrangian, similar to {\cite{Buchbinder:2007ak}}, which is local and does not contain higher-order derivative terms. Then, we derive a Lagrangian that allows for higher-order derivative terms, extending {\cite{Francia:2005bu}}, and can be expressed solely in terms of three fields.

\subsubsection{Unconstrained Form of Lagrangian for CS Theory without Higher-Order Derivatives}

Constraints on the fields have been removed up to this point. Next, we aim to eliminate traceless constraints on gauge parameter. The idea of this method involves introducing a compensator field {\cite{Francia:2002aa}}, $s_4$, with gauge transformation $\delta | s_4 \rangle = l_2 | \lambda \rangle$. By fixing gauge such that $s_4$ becomes zero, tracelessness of gauge parameter is induced. Of course, the Lagrangian with addition of $s_4$ must maintain gauge invariance, and EOM for new Lagrangian should be equivalent to those of the original $\mathcal{L}_1$(\ref{Lag_1f}) and $\mathcal{L}_4$(\ref{Lag_4f}). The introduction of $s_4$ is relatively straightforward in the case of $\mu = 0$. However, for non-zero $\mu$, the process becomes highly nontrivial. To address this challenge, we consider general situation that highlights the structure causing this problem and clarify how to incorporate new field $s_4$ into the Lagrangian (see Appendix \ref{compensator_general} for general approach). As a result, we obtain a gauge invariant Lagrangian that includes $s_4$, and both fields and gauge parameter within it remain unconstrained. It is written as follows:
\begin{gather}
 \mathcal{L}_5  =  \langle \overrightarrow{\phi_5} |
      \begin{pmatrix*}
        \partial^2  + L_X^{\dag} L_X & - L_X^{\dag} L_{X_1}^{\dag} & 0 &
        l_2^{\dagger} & 0\\
        - L_{X_1} L_X & - 2 \partial^2  + L_{X_2}^{\dag} L_{X_2} & 0 & 1 & -
        l_2^{\dag}\\
        0 & 0 & L_{(44)} & - \tilde{L}_{X_2} & {L'}^{\dag}_{X_2}\\
        l_2 & 1 & - \tilde{L}_{X_2}^{\dag} & 0 & 0\\
        0 & - l_2 & L'_{X_2} & 0 & 0
            \end{pmatrix*}
 | \overrightarrow{\phi_5} \rangle 
 \label{Lag_5f} \\
 \delta |\overrightarrow{\phi_5} \rangle  = 
     \begin{pmatrix*}
        \tilde{L}_X^{\dag}\\
        L'_X\\
        l_2\\
        - (\partial^2_{} + L_{X_2}^{\dag} L_{X_2}) \frac{\mu}{X \sqrt{2
        X_1}}\\
        - L_{X_3} L_{X_2}  \frac{\mu}{X \sqrt{2 X_1}}
   	\end{pmatrix*} | \lambda \rangle ,\qquad \qquad 
   	 | \overrightarrow{\phi_5} \rangle  
   	:=	\begin{pmatrix*}
   		|s_1 \rangle\\
   		|s_2 \rangle\\
   		|s_4 \rangle\\
   		|a_2 \rangle\\
   		|a_3 \rangle
   	\end{pmatrix*} \qquad\\
      \qquad \qquad \qquad
      L'_{X_k}  : =  \sqrt{X_{k + 1}} L_{X_k} \frac{1}{\sqrt{X_k}}  ,\quad 
      L_{(44)}  : =  - \frac{\mu}{X_2  \sqrt{2 X_3}}  (\partial^2 +
      L_{X_4}^{\dag} L_{X_4})  \frac{\mu}{X_2  \sqrt{2 X_3}} \quad.
\end{gather}

The equivalence between $\mathcal{L}_5$(\ref{Lag_5f}) and $\mathcal{L}_4$(\ref{Lag_4f}) can be understood as follows. First, by employing gauge transformation:
\begin{align}
  \delta |s_4 \rangle &= l_2 | \lambda \rangle\quad,
\end{align}
we gauge-fix $s_4$ to zero. This results in the gauge parameter having the constraint $l_2 | \lambda \rangle = 0$, and the Lagrangian reduces to $\mathcal{L}_4$. As the equivalence between $\mathcal{L}_4$ and $\mathcal{L}_1$ has already been mentioned, it ultimately proves that $\mathcal{L}_5$ is equivalent to $\mathcal{L}_1$. Therefore, this Lagrangian $\mathcal{L}_5$(\ref{Lag_5f}) describe CS fields, as deriving Wigner eqs.(\ref{kg_mod})-(\ref{trace_mod}) in Section \ref{wigner} via the Lagrangian $\mathcal{L}_1$(\ref{Lag_1f}) in Section \ref{lag_const}.

To make a comparison with {\cite{Buchbinder:2007ak}}, the field $C$ in {\cite{Buchbinder:2007ak}} needs to be eliminated through the EOM. This reveals its correspondence to the Lagrangian in $\mathcal{L}_5$ when $\mu \rightarrow 0.$\footnote{Our $s_1$, $s_2$, $s_4$, $a_2$, $a_3$, $\lambda$ respectively correspond to $\varphi$, $D$, $\alpha$, $\lambda_2$, $\lambda_3$, $\varepsilon$ in {\cite{Buchbinder:2007ak}}. To be more precise, our fields correspond to the sum over all helicities $n$ of the corresponding fields in {\cite{Buchbinder:2007ak}}, like $s_1 \sim \sum_n \varphi_n$, etc.
}
Alternatively, the correspondence can be seen by introducing a new field corresponding to $C$ in $\mathcal{L}_5$.\footnote{A new field $a_1$ corresponding to $C$ from ``triplet'' fields $(\varphi, C, D)$ could be introduced by further  EOM-equivalent modification.
}

This Lagrangian is local and has neither higher-order derivatives nor delta functions.\footnote{{\cite{Burdik:2019tzg}} obtained Lagrangians of different forms with these features. Please also refer to the footnote in the Introduction.
}
Consequently, we have obtained a conventional form of the CS Lagrangian in field theory. As a result, this Lagrangian $\mathcal{L}_5$ offers significant advantages in terms of practicality and facilitates various standard calculations within field theory. Therefore, we believe that this Lagrangian $\mathcal{L}_5$ possesses considerable utility.

The most interesting points about obtaining this Lagrangian $\mathcal{L}_5$ are as follows:
First, let's remind ourselves of the case of HS theories (i.e., $\mu = 0$). In the tensionless limit of string theory, the Virasoro algebra is simplified. The BRST charge is then written in terms of HS operators: $\partial^2, l_1, l_1^{\dag}$, and the resulting Lagrangian is expressed using fields called "triplets." In contrast, when \ theories are formulated using BRST formulation, the HS operators include $l_2, l_2^{\dag}, N$ as well as $\partial^2, l_1, l_1^{\dag}$. Nevertheless, the Lagrangian can still be constructed in a similar fashion, allowing us to identify ``triplet'' fields within it. Furthermore, Lagrangian {\cite{Buchbinder:2007ak}} relates to the BRST formulated Lagrangian through partial gauge fixing and partial use of the EOM. Notably, in {\cite{Buchbinder:2007ak}}, the triplet fields written as ($\varphi, C, D$) appear explicitly in the Lagrangian, revealing an interesting relationship with tensionless string theory.

Intriguingly, if this relationship for HS extends to the case of CS theories (i.e., $\mu \neq 0$), it might offer insights into the connection between string theory and CS theories. However, a BRST formulation of the Lagrangian for CS theories (with vector index field) remains elusive. This presents a challenge for future research. For more information on the relationship between CS theories and string theory, please refer to {\cite{Savvidy:2003fx,Mourad:2004fg,Mourad:2005rt}}.

\subsubsection{Unconstrained Form of Lagrangian for CS Theory with Third-Order Derivatives}

In {\cite{Francia:2005bu}}, for $\mu = 0$, an unconstrained Lagrangian is assumed that can be expressed in terms of a single field and includes non-local terms. By introducing compensator and auxiliary fields, a Lagrangian without non-local terms is derived. In our case, to obtain a similar Lagrangian for $\mu \neq 0$, we start from the Lagrangian $\mathcal{L}_5$(\ref{Lag_5f}) and eliminate $s_2$ and $a_2$ themselves from the Lagrangian using their EOM, while keeping the compensator field $s_4$.

By using the EOM for $\langle a_2 | $ and $\langle s_2 | $, we can solve $|s_2 \rangle$ and $|a_2 \rangle$, respectively, as follows:
\begin{alignat}{2}
    \langle a_2 |  & :\quad |s_2 \rangle && =  - l_2 |s_1 \rangle + \tilde{L}_{X_2}^{\dag} |s_4 \rangle\\
    \langle s_2 |  & : \quad |a_2 \rangle && =  L_{X_1} L_X |s_1 \rangle + (2 \partial^2  - L_{X_2}^{\dag} L_{X_2}) |s_2 \rangle + l_2^{\dag} |a_3 \rangle \quad .
\end{alignat}
Using these solutions and their conjugates, we can eliminate $s_2$ and $a_2$ from the Lagrangian as follows:\footnote{Here, the operator $E$ is the one that appears when the Lagrangian $\mathcal{L}_1$(\ref{Lag_1f}) is written as: $ \mathcal{L}_1 = \langle s_1 |E | s_1 \rangle$. See also Appendix \ref{convention}.
}
\begin{gather}
        \mathcal{L}_3=  \langle \overrightarrow{\phi_3}|
        \begin{pmatrix*}
       		E & L_{(14)} & l_2^{\dagger 2}\\
       		 L_{(14)}^{\dag} & 
          	L_{(44)} - \tilde{L}_{X_2} (2 \partial^2  - L_{X_2}^{\dag} L_{X_2})
          	\tilde{L}_{X_2}^{\dag}
        	& - \tilde{L}_{X_2} l_2^{\dag}         +{L'}^{\dag}_{X_2}\\
        	l_2^2 & - l_2 \tilde{L}_{X_2}^{\dag} + L'_{X_2} & 0
        \end{pmatrix*}
 |\overrightarrow{\phi_3} \rangle
  \label{Lag_3f}\\
\delta | \overrightarrow{\phi_3} \rangle  = 
        \begin{pmatrix*}
      	  \tilde{L}_X^{\dag}\\
      	  l_2\\
       	- L_{X_3} L_{X_2}  \frac{\mu}{X \sqrt{2 X_1}}
        \end{pmatrix*}
      | \lambda \rangle ,\qquad\qquad\qquad\qquad
      | \overrightarrow{\phi_3} \rangle  : = \begin{pmatrix*}
      	|s_1 \rangle\\
      	|s_4 \rangle\\
      	|a_3 \rangle
      \end{pmatrix*}\quad\qquad\\
   \qquad \qquad
   \begin{aligned}
     & E && =  (1 - l_2^{\dagger} l_2) \partial^2  \;+
		L_{\text{(de Donder)}}^\dag 
		L_{\text{(de Donder)}} \;+
   		 \left\{ l_2^{\dag} \frac{\mu}{X_1 \sqrt{2 X_2}}
      L_{X_3}^{\dag} l_2^2  + h.c. \right\}\\
    &  L_{(14)} && :=  (2 l_2^{\dagger} \partial^2  - L_X^{\dag} L_{X_1}^{\dag} - l_2^{\dagger} L_{X_2}^{\dag} L_{X_2}) \tilde{L}_{X_2}^{\dag}\quad.
    \end{aligned}
\end{gather}

We have obtained a Lagrangian that can be expressed in terms of three fields, similar to the formulation by Francia and Sagnotti {\cite{Francia:2005bu}}, even in the presence of $\mu$.

By eliminating the compensator field $s_4$ through its gauge transformation, and eliminating $a_3$ by the EOM using the procedure in Appendix \ref{eom_unconstl} after eq.(\ref{EOM_2f}), Lagrangian $\mathcal{L}_3$(\ref{Lag_3f}) reduces to the Fronsdal-like Lagrangian $\mathcal{L}_1$(\ref{Lag_1f}), as expected. Therefore, this Lagrangian $\mathcal{L}_3$ is confirmed to describe CS fields, as deriving Wigner equations.

A point to note is that the fields in this Lagrangian do not directly correspond to those in {\cite{Francia:2005bu}}. In fact, while our field $a_3$ is gauge invariant in case of $\mu \rightarrow 0$, the corresponding field $\beta$ in {\cite{Francia:2005bu}} is not. On the other hand, $\lambda_3$ in {\cite{Buchbinder:2007ak}} is gauge invariant and directly corresponds to our field $a_3$. This relationship is also discussed in {\cite{Buchbinder:2007ak}}, where it is explained that these models are related by a field redefinition. Consequently, we believe that Lagrangian $\mathcal{L}_3$(\ref{Lag_3f}) shares essentially the same properties as that in {\cite{Francia:2005bu}}.

It is interesting that if we consider the EOM while keeping the compensator field $s_4$, as was done for the case of HS in {\cite{Francia:2005bu,Francia:2006hp}}, there is a possibility of obtaining equations for the compensator system for the CS case. As considered for the case of HS theories in {\cite{Francia:2005bu,Francia:2006hp}}, the compensator system is related to a geometric form. Therefore, one of the main reasons why Lagrangian $\mathcal{L}_3$ is a theoretically intriguing result is that our Lagrangian opens up the possibility of a path towards a CS version (i.e., $\mu \neq 0$) of the geometric form.

As a further investigation of $\mathcal{L}_3$, we propose extending the approach of {\cite{Francia:2007qt}}, which explored HS fields interacting with external currents using local and non-local formulations, to the case of CS fields. The concluding section \ref{summary} discusses this future direction of CS field interactions in more detail.

\section{Conclusion and Perspective}\label{summary}

\subsection{Conclusion }

In this paper, we have considered a free bosonic CS field described by totally symmetric tensor fields of all ranks in arbitrary dimensional flat spacetime. 

In Section \ref{wigner}, we considered the Wigner eqs.(\ref{kg})-(\ref{trace}) to impose the requirement that the field obeys the irreducible CS representation. Performing a Fourier-like transformation (\ref{fouriertr}), we re-expressed the equations in terms of Fock state vector, generated by auxiliary creation operators, and HS operators acting on it. We then introduced trace generating operator and modified the Wigner equations to include a traceless condition, resulting in eqs.(\ref{kg_mod})-(\ref{trace_mod}).

In Section \ref{lag_const}, we derived the Lagrangian that leads to the modified Wigner eqs.(\ref{kg_mod})-(\ref{trace_mod}). We achieved this by rewriting the Lagrangian $\mathcal{L}_s$(\ref{Lag_spin}) previously obtained for fields with spinor indices in four-dimensional spacetime using the BRST formulation. This resulted in a Lagrangian $\mathcal{L}_1$(\ref{Lag_1f}) written in terms of fields with vector indices. This process involved a systematic rewrite using a unitary-like operator (\ref{unitaryop}) and extension to arbitrary $D$ dimensions. The resulting Lagrangian $\mathcal{L}_1$ is gauge invariant and described by a single double-traceless field. We confirmed that this Lagrangian $\mathcal{L}_1$ leads to the modified Wigner eqs.(\ref{kg_mod})-(\ref{trace_mod}) given in Section \ref{wigner} under the de Donder-type gauge fixing eq.(\ref{gauge_fixing_dtrace}) with tracelessness, or equivalently eqs.(\ref{gauge_fixing1}) and (\ref{gauge_fixing2}). 

Having confirmed that the Lagrangian describes a CS theory, we proceeded to Section \ref{lag_unconst}. Here, we derived unconstrained Lagrangians  $\mathcal{L}_5$(\ref{Lag_5f}) and $\mathcal{L}_3$(\ref{Lag_3f}) by performing equivalent modifications on the Lagrangian $\mathcal{L}_1$ from Section \ref{lag_const}. These unconstrained Lagrangians have no constraints on either the fields or the gauge parameter. To achieve an unconstrained Lagrangian, we introduced auxiliary fields $s_2$, $a_2$, $a_3$ and a compensator $s_4$. The Lagrangian derivation employed a general method that incorporates CS parameter. This results in an unconstrained Lagrangian $\mathcal{L}_5$ with five fields and one gauge parameter, which naturally extends the quartet formulation for $\mu = 0 $ {\cite{Buchbinder:2007ak}}. Notably, this Lagrangian $\mathcal{L}_5$ avoids non-local terms, delta functions, and higher-order derivative terms -- hallmarks of conventional field theory Lagrangians.

We also derived another unconstrained Lagrangian $\mathcal{L}_3$ with three fields including a compensator $s_4$ and an auxiliary field $a_3$, by allowing terms up to the third-order derivative. This is similar to the {\cite{Francia:2005bu}} in the geometric approach for $\mu = 0$ and has been obtained for the first time for $\mu \neq 0$.

We believe that these results provide a significant foundation for further exploring CS field theories.

\begin{figure}[H]
	\centering
    \includegraphics[width=1.00\hsize]{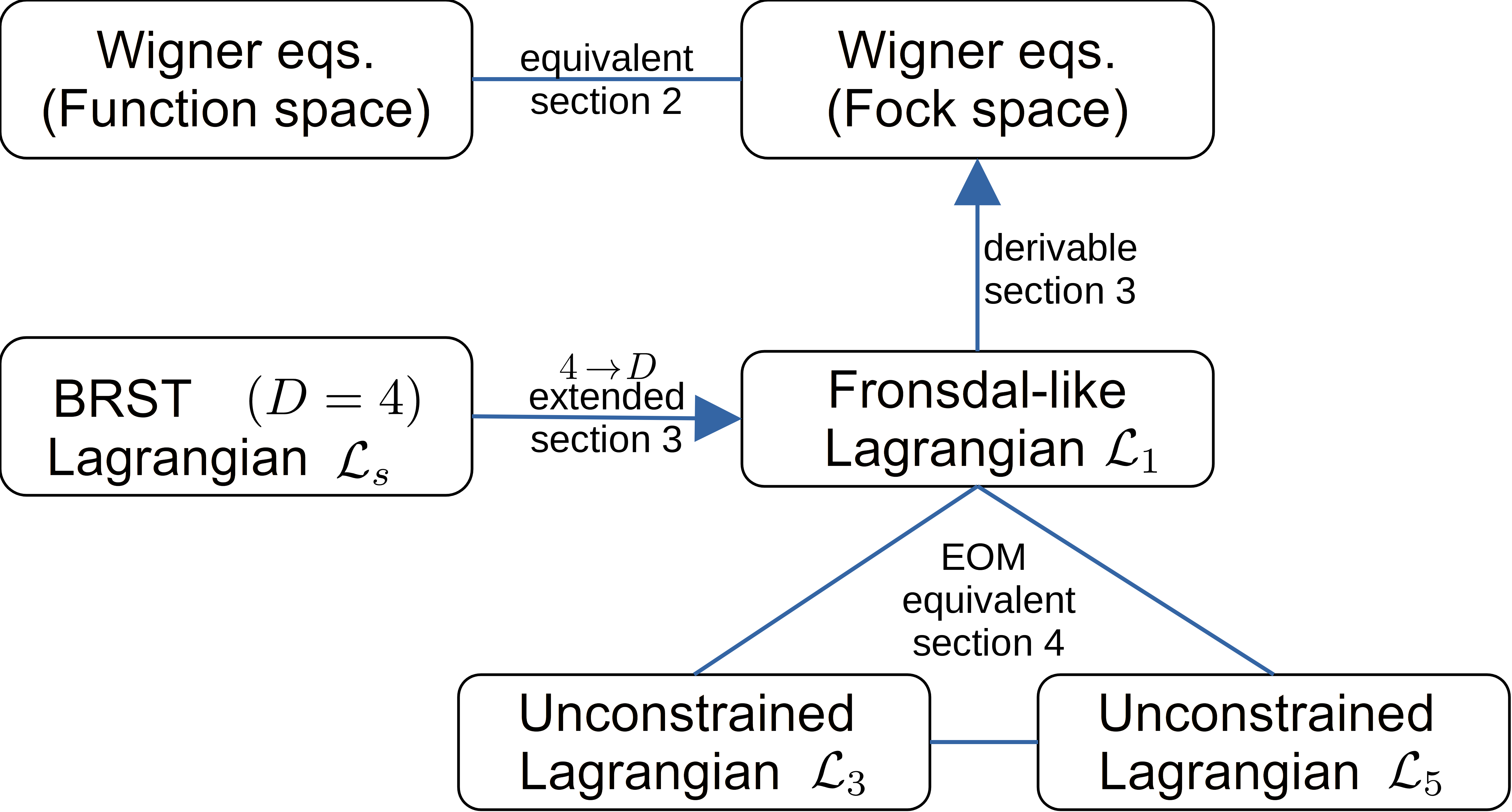}
  	\caption[Diagramatical summary]{\label{fig2}Diagramatical summary of paper for CS theory}
\end{figure}

\subsubsection*{Conclusion in Short (see Figure \ref{fig2})}
  \paragraph*{\quad Novel Lagrangians $\mathcal{L}_5$ and $\mathcal{L}_3$ for CS Fields:}
	\begin{itemize}
  		\item No constraints on fields or gauge parameter.
  		\item No non-local terms or delta functions. 
  		\item $\mathcal{L}_5$: Five fields, one gauge parameter ($\mu \neq 0$): Extension of Buchbinder et al. {\cite{Buchbinder:2007ak}}   (no higher-order derivatives).
  		\item $\mathcal{L}_3$: Three fields, one gauge parameter ($\mu \neq 0$): Extension of Francia and Sagnotti {\cite{Francia:2005bu}}   (up to third-order derivatives).
	\end{itemize}
  \paragraph*{\quad Verification:}
	\begin{itemize}
  		\item Lagrangians $\mathcal{L}_5$ and $\mathcal{L}_3$ lead to the Wigner equations.
  		\item Consistent with previous literature.
	\end{itemize}
\subsection{Perspective}

The Lagrangians obtained in this paper provide a significant stepping stone for future research in CS theory. In particular, the Lagrangian $\mathcal{L}_5$(\ref{Lag_5f}) stands as the closest form yet to the unknown BRST-formulated Lagrangian, having interesting relation to string theory, and Lagrangian $\mathcal{L}_3$(\ref{Lag_3f}) serves as a valuable foundation for a geometrical understanding of CS fields. These results open up several exciting avenues for further exploration, which we outline below.

\subsubsection*{BRST Formulation}

The BRST formulation offers a powerful approach to connect spacetime symmetries to Lagrangians, potentially deepening our understanding of CS theories and their connection to string theory (Ref. {\cite{Sagnotti:2003qa}}). While formulating a $\mu \neq 0$ BRST Lagrangian remains elusive, it could lead to advancements in:
\begin{itemize}
  \item Understanding spacetime symmetries through Lagrangian encoding. 
  \item Uncovering new insights into string theory.
  \item Simplifying Lagrangian construction for CS theories.
\end{itemize}
  \subparagraph*{Difficulty and Idea:} 
  Here, we discuss the difficulties encountered when attempting to construct a BRST Lagrangian for CS theories (using vector index fields) and present an idea for a potential solution. We consider an algebra with constraints given by the Wigner eqs.(\ref{kg_hs})-(\ref{trace_hs}) or their modified forms eqs.(\ref{kg_mod})-(\ref{trace_mod}), which serve as conditions for the irreducible representation. An attempt to use the BRST method with a closed algebra without second-class constraints has already been discussed in {\cite{Bengtsson:2013vra}}. However, we will consider it in a similar form to this paper and {\cite{Buchbinder:2005ua}}. As demonstrated in this  study, one of the four Wigner eqs., e.g. eq.(\ref{gauge_mod}), can be derived from the gauge transformation form inherent in the Lagrangian $\mathcal{L}_1$(\ref{Lag_1f}). Therefore, we tentatively choose to ignore these conditions.\footnote{In
  {\cite{Bengtsson:2013vra,Burdik:2019tzg}}, they consider the algebra including the remaining constraint conditions too.
  }
  This leaves us with   three remaining constraints. By performing a suitable number operator dependent normalization change for the field, we can rewrite these three constraints in a form that does not involve the number operator:
  \begin{align}
      \partial^2 | \phi'' \rangle & =  0\\
      (l_1  - \mu) | \phi'' \rangle & =  0\\
      l_2 | \phi'' \rangle & =  0 \quad .
  \end{align}
  Considering the operators appearing here and their Hermitian conjugates, we find that the following seven operators form a closed algebra:
  \begin{alignat}{7}
      \partial^2 & ,\quad  \mu & ,\quad L_1  :=  l_1 - \mu & ,\quad  L_1^{\dag}  =  l_1^{\dag} - \mu & , \quad l_2 & ,\quad  l_2^{\dag} & ,\quad  X_1  =  N + D / 2.
  \end{alignat}
  However, $\mu$ and $X_1$ cannot be regarded as constraints. This means that second-class constraints appear, leading to difficulties. One way to solve this problem is to introduce new creation and annihilation operators and consider a new representation. In fact, a similar situation arises in the   case of massive HS algebras, and a specific solution for this case is known {\cite{Buchbinder:2005ua}}. It would be very interesting if a similar approach could be used for the CS case. However, in this case, as in {\cite{Bengtsson:2013vra}}, there may exist a problem that two self-conjugate constraints appear, which must be solved.
  
  Another perspective on the connection between CS theory and string theory is presented in {\cite{Savvidy:2003fx,Mourad:2004fg,Mourad:2005rt}}.

\subsubsection*{Geometrical picture of CS field}

It is possible to obtain equations for a compensator system in the CS case ($\mu \neq 0$) by retaining the compensator field $s_4$ in the EOM. This approach aligns with the connection between the compensator system and a geometric form observed in HS theories {\cite{Francia:2005bu,Francia:2006hp}}. Therefore, Lagrangian $\mathcal{L}_3$(\ref{Lag_3f}) may be a key to unlocking a CS version of the geometric form.

\subsubsection*{Interaction in CS theories}

While this paper has concentrated on the free field theory of CS, the interaction of these fields remains a central theme in this area of research. Notably, prior studies, such as those in {\cite{Metsaev:2017cuz,Bekaert:2017xin}}, have investigated the cubic interaction. Moreover, extensive research exists on the interaction of HS fields. The methods employed in these HS interaction studies, as comprehensively summarized in the conclusion section of {\cite{Metsaev:2017cuz}}, hold potential for extension to CS. In particular, the unconstrained formulation may be well-suited for considering interactions, potentially along two interesting directions:
\begin{itemize}
  \item Interaction via conserved current
  
  As noted in {\cite{Francia:2007qt}}, one of the attractive features of the unconstrained formulation is the ability of the field to couple to conserved currents. The authors in {\cite{Francia:2007qt}} employed the local and non-local formulations for HS fields, presented in {\cite{Francia:2002aa,Francia:2005bu}}, to discuss the HS fields interacting with external currents. In {\cite{Bekaert:2009ud}}, the authors studied the cubic coupling between scalar fields and a HS field through the Noether interaction with the conserved current. Building on these approaches, {\cite{Bekaert:2017xin}} extended the technique to study the interaction of a CS field with two scalar fields based on the Schuster-Toro Lagrangian {\cite{Schuster:2014hca}}. Since our Lagrangian $\mathcal{L}_3$(\ref{Lag_3f}) bears a strong resemblance to the one in {\cite{Francia:2005bu}}, a more direct extension of the work presented in {\cite{Francia:2007qt}} and {\cite{Bekaert:2009ud}} to the CS case might be possible.
  
  \item BRST and BRST-BV
  
  The BRST formulation allows for a clear description of gauge invariant cubic interactions due to the BRST charge. For example, {\cite{Metsaev:2012uy}} considers cubic interactions in HS theory using the BRST-BV approach, and an application of this technique to CS would be highly interesting.
\end{itemize}

\subsubsection*{Other Potential Research Themes}

In addition to the above, the method for deriving unconstrained form of CS Lagrangians $\mathcal{L}_5$(\ref{Lag_5f}) and $\mathcal{L}_3$(\ref{Lag_3f}) in this paper could be applied in the following contexts:
\begin{itemize}
  \item AdS spacetime
  \item Fermionic  
  \item Supersymmetric  
  \item Mixed symmetry
\end{itemize}

\section*{Acknowledgments}
  The author would like to thank I.L. Buchbinder and V.A. Krykhtin for their helpful discussions and valuable comments. The author would also like to thank T. Morozumi for general discussions and arranging the hospitality during the author's visit to the theoretical particle physics group at Hiroshima University.
  
\appendix\section{Conventions and Definitions}\label{convention}
\begin{itemize}
  \item Definition of Hermite Conjugation $\dag$ in Vector Index:
  \begin{alignat}{3}
      i  & \xleftrightarrow[\dag]{}  - i & ,\quad  \partial_{\mu} &  \xleftrightarrow[\dag]{}  - \partial_{\mu} & ,\quad  {a^{\mu}}  &      \xleftrightarrow[\dag]{} a^{\dag \mu} \quad .
  \end{alignat}
  \item HS Operators and Metric in Vector Index: 
  \begin{alignat}{5}
        [a^{\mu}, a^{\dag \nu}]&=\eta_{\mu \nu}&&,\qquad  l_1 && :=  a \cdot \partial &&, \qquad l_1^{\dag} && =  - a^{\dag} \cdot \partial 
\\
       &=  \text{diag} (-, +, +,\cdots)  &&,\qquad  l_2 && :=  \frac{1}{2} a^2 &&,  \qquad l_2^\dag && = \frac{1}{2} a^{\dag 2}
\\
      X_m&:= X+m ,m \in \mathbb{Z} && ,\qquad X && := N + D / 2- 1 
        &&, \quad\;\;  N && :=  a^{\dag} \cdot a
\\ 
       [l_1, l_2^{\dag}] & =  - l_1^{\dag} &&,\quad l_1 N && =  (N + 1) l_1  &&,\quad N l_1^{\dag} && =  l_1^{\dag} (N + 1)
\\
       [l_2, l_1^{\dag}] & =  - l_1   &&,\quad  l_2 N && =  (N + 2) l_2  &&,\quad N l_2^{\dag} && =  l_2^{\dag} (N + 2)
\\
        [l_1, l_1^{\dag}] & =  - \partial^2&& ,\; [l_2, l_2^{\dag}] && =  X + 1  &&.\qquad \;\;\;&&
  \end{alignat}
  Here, $\mu = 0, 1, \ldots D - 1$ and ``dot'' indicates contraction of two vectors by metric: $a \cdot b :=  a_{\mu} \eta^{\mu \nu} b_{\nu}$. We have defined shifted number operator $X$ for more compact expressions throughout the paper. For totally symmetric HS fields, $l_1 | \phi \rangle = 0$ and $l_2 | \phi \rangle = 0$ are equivalent to the divergence free equation ($\partial^{\mu_1} \phi_{\mu_1 \cdots} (x) = 0$) and the trace free equation ($\eta^{\mu_1 \mu_2} \phi_{\mu_1 \mu_2 \cdots} (x) = 0$), respectively. The relationship $| \phi \rangle = \sum_{n = 0}^{\infty} a^{\dag \mu_1} \cdots a^{\dag \mu_n} | 0 \rangle \phi_{\mu_1 \ldots \mu_n} (x)$ shows this equivalence. See {\cite{Buchbinder:2005ua,Buchbinder:2006nu}} for example. 
  
  \item Various Operators used in Main Text
	\begin{enumerate}
	 \item $n$-trace projection operator $P_n$ satisfying $l_2^n P_n | \text{any} \rangle = 0$, is commonly expressed recursively:
      \begin{alignat}{2}
          P_n & :=  \Pi_n P_{n + 1} & ,  \quad \Pi_n & :=  1 - l_2^{\dag n}
          \frac{1}{n! (X_n)^{\underline{n}}} l_2^n\quad .
      \label{proj}
      \end{alignat}     
      Here, $\Pi_n$ is understood as a projection operator onto the $n$-traceless fields within $n + 1$ traceless Fock space. See Appendix \ref{factorial} for definition of the factorial power symbol $(x)^{\underline{n}}$.
	 \item Generalized divergence operators that reduce to $l_1$ as $\mu \rightarrow 0$ are as follows:
  \begin{alignat}{3}
           L_{X_k} & =  l_1 - \frac{\mu}{\sqrt{2 X_k}} & ,  \quad L'_{X_k} & = 
          \sqrt{X_{k + 1}} L_{X_k} \frac{1}{\sqrt{X_k}} &,\quad \tilde{L}_{X_k} & =  L_{X_k} + \frac{\mu}{X_k \sqrt{2 X_{k + 1}}} l_2 \;.
       \label{div_op}
  \end{alignat}
  \item De Donder operator used in gauge fixing of eq.(\ref{gauge_fixing_dtrace}) is as follows:
\begin{align}
 	L_{\text{(de Donder)}} 
  		&=L_X + \tilde{L}_X^{\dag} l_2\\
  		&\equiv l_1 + l_1^{\dag} l_2 + \mu \left( \frac{1}{\sqrt{2 X}}
  							\Pi_1 + \frac{1}{\sqrt{2 X_1}} l_2 \right) \quad .
\end{align}
  
  \item Operator $E$ defined through Fronsdal-like Lagrangian $\mathcal{L}_1$(\ref{Lag_1f}) and appeared also in unconstrained Lagrangian   $\mathcal{L}_3$(\ref{Lag_3f}) is as follows:
  
  \begin{align}  
  	E & :=\begin{pmatrix}
  		1 & l^{\dag}_2
  	\end{pmatrix} 
  	\begin{pmatrix*}[c]
  		\partial^2  + L_X^{\dag} L_X & L_X^{\dag} L_{X_1}^{\dag}\\
  		L_{X_1} L_X & - 2 \partial^2  + L_{X_2}^{\dag} L_{X_2}
  	\end{pmatrix*} 
  	\begin{pmatrix}
  		1\\
  		l_2
  	\end{pmatrix} \notag\\
  	 & \equiv  (1 - 2 l^{\dag}_2 l_2) \partial^2 + L_X^{\dag} (1 + l_2^{\dag} l_2 ) L_X + l_2^{\dag} L_{X_1}L_X + L_X^{\dag} L^{\dag}_{X_1} l_2 \notag \\	
  	 & \equiv(1 - l_2^{\dagger} l_2) \partial^2  \;+
  	 L_{\text{(de Donder)}}^\dag L_{\text{(de Donder)}} \;+
  	 \left\{ l_2^{\dag} \frac{\mu}{X_1 \sqrt{2 X_2}}
  	 L_{X_3}^{\dag} l_2^2  + h.c. \right\}\quad.
  	 \label{EOMop_formula}  	 
  \end{align}  
  \item Other operators appeared in unconstrained Lagrangians are as follows:
  \begin{align}
          L_{(14)} & =  (2 l_2^{\dagger} \partial^2  - L_X^{\dag}
        L_{X_1}^{\dag}  - l_2^{\dagger} L_{X_2}^{\dag}
        L_{X_2}) \tilde{L}_{X_2}^{\dag}\\
        L_{(44)} & =  - \frac{\mu}{X_2  \sqrt{2 X_3}}  (\partial^2 +
        L_{X_4}^{\dag} L_{X_4})  \frac{\mu}{X_2  \sqrt{2 X_3}}\quad .
   \end{align}     
\end{enumerate}
  \item HS Operators and Hermite Conjugation $\dag$ in Spinor Index($a,b=1,2$):
  \begin{alignat}{5}
      i  & \xleftrightarrow[\dag]{}  - i & , \quad \partial_{a \dot{b}} &  \xleftrightarrow[\dag]{}  - \partial_{b \dot{a}} & , \quad a &  \xleftrightarrow[\dag]{}  \bar{c} & , \quad \bar{a} &  \xleftrightarrow[\dag]{}  c
  \end{alignat}
  \begin{alignat}{5}
        [\bar{a}^{\dot{a}}, \bar{c}_{\dot{b}}] & = \delta^{\dot{a}}_{\dot{b}} ,\; N_{\bar{a}}:=\bar{c}_{\dot{a}}\bar{a}^{\dot{a}}&&,&    l_1^s & =  \partial_{a \dot{b}} a^a  \bar{a}^{\dot{b}}  \qquad&&,&  L_1^{s 	\dag} & =  l_1^{s \dag} - \mu
\\
               [a_a, c^b] & =  \delta_a^b ,\; N_a:=c^aa_a &&,&  {l_1^{\dag}}^s & = - c^b \partial_{b \dot{a}}  \bar{c}^{\dot{a}}   \qquad&&,&    L^s_1 & =  l^s_1 - \mu
\\
         N^s & =  \frac{1}{2} (N_a +N_{\bar{a}}) \;\;&&,& [ {l^s_1}^{\dag}, l^s_1 ] & =  2 (N^s + 1) \partial^2 &&,&  l^s_1 N^s & =   (N^s + 1) l^s_1
\\        
        \sigma^{\mu}_{a \dot{b} } \bar{\sigma}_{\mu}^{ \dot{a} b}  & =  - 2 \delta^b_a \delta^{\dot{a}}_{\dot{b}} \qquad\;\;&&,&   \bar{\sigma}_{\mu} \sigma_{\nu} + \mu \leftrightarrow \nu & =  - 2 \eta_{\mu \nu}\qquad&&,& N^s l_1^{s \dag} & =  l_1^{s \dag} (N^s + 1)
\\        
        \bar{\sigma}_{\mu}^{ \dot{a} b} \sigma^{\mu}_{a \dot{b} } & =  - 2    \delta^b_a \delta^{\dot{a}}_{\dot{b}} \qquad\;\;&&,&  \sigma_{\nu}  \bar{\sigma}_{\mu}  + \mu \leftrightarrow \nu & =  - 2 \eta_{\mu  \nu}\qquad&&,& (\bar{\sigma}_{\mu} )^{\dot{b} a}  (\sigma_{\nu})_{a  \dot{b}} & =  - 2 \eta_{\mu \nu} \qquad .
  \end{alignat}
\end{itemize}

\section{Fourier-like Transformation between Real Variables and Creation Operators}\label{fourier}

First, we confirm that the Fourier transformation is typically written as follows:
\begin{alignat}{2}
    \psi (w) & :=  \int \frac{d^D \xi}{(2 \pi)^{D / 2}} e^{- i w \cdot \xi}
    \Psi (\xi) & ,\quad  \Psi (\xi) & =  \int \frac{d^D w}{(2 \pi)^{D / 2}} e^{i
    w \cdot \xi} \psi (w) \quad .
\end{alignat}

Now, let's consider a transformation similar to the usual Fourier transformation, but using the creation operator $a^{\dag}$ instead of $w$. We define $\psi (a^{\dag})$ as the replacement of $w$ with $a^{\dag}$ in the function $\psi (w)$. Then, we correspond this to a Fock state vector as $| \psi \rangle   :=  \psi (a^\dag) | 0 \rangle $.

Now, consider a Fourier-like transformation between the function $\Psi (\xi)$ of the real variables $\xi^{\mu}$ and the Fock state vector $| \psi \rangle$, generated by creation operators $a^{\dag}$, as follows:
\begin{alignat}{2}
    \psi (a^{\dag}) & :=  \int \frac{d^D \xi}{(2 \pi)^{D / 2}} e^{- i
    a^{\dag} \cdot \xi} \Psi (\xi) &  &   \\
    \Psi (\xi) & =  (2 \pi)^{D / 2} \langle 0 | \delta (\xi - i a) | \psi
    \rangle    & , \quad | \psi \rangle  & :=  \psi (a^{\dag}) | 0 \rangle 
 \quad .
\end{alignat}
Duality $\Psi (\xi)  \leftrightarrow  | \psi \rangle$ needs to be verified, which can be done using the properties of coherent states.\footnote{The coherent state $|q \rangle$ is defined for complex numbers $p_{\nu}$ and $q_{\nu}$ with the following properties(see also Appendix \ref{coherent}):
\begin{align}
	| q \rangle := e^{a^{\dag}  \cdot q} | 0 \rangle & , & a_{\nu} | q  \rangle = q_{\nu} | q \rangle & , & \langle p |   = \langle 0 |
	e^{a  \cdot p^{\ast}}   & , & \langle p | a_{\nu}^{\dag}   =  \langle p | p_{\nu}^{\ast}   & , & \langle p | q \rangle   =  e^{p^{\ast} \cdot q}\;.
\end{align} 
}
For a coherent state with a complex vector in $D$-dimensions, one can express it as follows: $e^{- i a^{\dag} \cdot \xi} | 0 \rangle  =  |- i \xi \rangle  ,  \langle  0| e^{a  \cdot \eta}  =  \langle \eta |  $, where $\xi$ and $\eta$ are real. Using these notations and properties, we find that the correct duality relation: 
\begin{table}[H]  
\centering 
\begin{tabularx}{1.0\textwidth}{X|X}	
\begin{eqnarray}
	&& (2 \pi)^{D / 2} \langle 0 | \delta (\xi - i a) {\color{blue}{\psi (a^{\dag})}} | 0 \rangle 
\notag\\
	&=&  (2 \pi)^{D / 2} \langle 0| \delta (\xi - i a) 
\notag\\ 
	&&   \times  {\color{blue}{\int \frac{d^D \xi'}{(2 \pi)^{D / 2}} e^{- ia^{\dag} \cdot \xi'} \Psi (\xi') }}| 0 \rangle 
\notag\\
	&=&  \int d^D \xi' \langle 0| \delta (\xi - i a) | - i \xi' \rangle \Psi (\xi')
\notag\\
	&=&  \int d^D \xi' \langle 0 | - i \xi' \rangle | \delta (\xi - \xi') \Psi (\xi') 
\notag\\
	&=&  {\color{red}{\Psi (\xi)}} \quad,\label{duality1}
\end{eqnarray} 
& 
\begin{eqnarray}
	&& \int \frac{d^D \xi}{(2 \pi)^{D / 2}} e^{- i a^{\dag} \cdot \xi} {\color{red}{\Psi (\xi)}} 
\notag\\  
	&=& \int \frac{d^D \xi}{(2 \pi)^{D / 2}} e^{- i a^{\dag} \cdot \xi} 
		{\color{red}{(2 \pi)^{D / 2}}} 
\notag\\
	&&   \times {\color{red}{\langle 0| \delta (\xi - i a) \psi (a^{\dag}) | 0 \rangle}}     
\notag\\
	&= & \int \frac{d^D \xi}{(2 \pi)^{D / 2}} e^{- i a^{\dag} \cdot \xi} (2 \pi)^{D / 2} 
\notag\\
	&&   \times \langle 0| \int \frac{d^D \eta}{(2 \pi)^D} e^{i (\xi - i a) \cdot \eta} \psi (a^{\dag}) | 0 \rangle 
\notag\\
	&= & \int \frac{d^D \eta d^D \xi}{(2 \pi)^D} e^{i (\eta - a^{\dag}) \cdot \xi} \langle 0| e^{a  \cdot \eta} \psi (a^{\dag}) | 0 \rangle 
\notag\\
	&= & \int d^D \eta \delta (\eta - a^{\dag}) \langle \eta | \psi  (a^{\dag}) | 0 \rangle   
\notag\\
	&= & \int d^D \eta \delta (\eta - a^{\dag}) \psi (\eta ) \langle \eta |  0 \rangle 
\notag\\
	&= & {\color{blue}{\psi ( a^{\dag}  ) }}\quad . \label{duality2}
\end{eqnarray}
\end{tabularx}
	\caption[Duality of Fourier-like transformation]{\label{table2}Duality relation for the Fourier-like transformation}
\end{table}
We can also find the following properties with any function $f$, similar to the standard Fourier transformation (assuming total derivative vanishes):
\begin{alignat}{3}
    f (i a)  | \psi \rangle  & =  \int\frac{d^D \xi}{(2 \pi)^{D / 2}} | - i \xi \rangle f (\xi)  \Psi (\xi)  
    &&,\quad  f (\xi) \Psi (\xi) && =  (2 \pi)^{D / 2} \langle 0|  \delta (\xi - i a) f (i a) | \psi \rangle
\\
    f (i a^{\dag})   | \psi \rangle & = \int \frac{d^D \xi}{(2 \pi)^{D / 2}} | - i \xi \rangle f (\partial_{\xi}) \Psi (\xi) 
    &&,\quad  f (\partial_{\xi}) \Psi (\xi) && =  (2 \pi)^{D / 2} \langle 0 | \delta (\xi - i a)  f (i a^{\dag})  | \psi  \rangle  \quad .
\end{alignat}

Thus, we find a simple duality($\leftrightarrow$) correspondence:
\begin{alignat}{3}
     \Psi (\xi) & \leftrightarrow  | \psi \rangle &\quad \Rightarrow \quad
          f (\xi_{\mu}) \Psi (\xi) & \leftrightarrow  f (i a) | \psi \rangle & ,\quad
      f (\partial_{\xi}) \Psi (\xi) & \leftrightarrow  f (i a^{\dag}) |
       \psi \rangle \quad .
       \end{alignat}
By further Fourier transforming from momentum to coordinate, we find correspondence:
\begin{alignat}{3}
    \Psi (p) & \leftrightarrow  \psi (x) & , \quad p \Psi (p) & \leftrightarrow 
    i \partial_x \psi (x) & , \quad \partial_p \Psi (p) & \leftrightarrow  i x
    \psi (x) \quad .
\end{alignat}
Now, we can rewrite Wigner equations using Fourier(-like) transformation $\Psi (p, \xi) \rightarrow | \tilde{\psi} (p) \rangle \rightarrow | \psi (x) \rangle$ \ as follows:
\begin{align}
\begin{aligned}
     \partial^2 \Psi (p, \xi) & =  0\\
     (p \cdot \partial_{\xi} - i \mu) \Psi (p, \xi) & =  0\\
     (p . \xi) \Psi (p, \xi) & =  0\\
     (\xi^2 - 1) \Psi (p, \xi) & =  0
\end{aligned} 
\quad&\rightarrow\quad
\begin{aligned}
     p^2 | \tilde{\psi} (p) \rangle & =  0\\
     (p \cdot a^{\dag} - \mu) | \tilde{\psi} (p) \rangle & =  0\\
     (p . a) | \tilde{\psi} (p) \rangle & =  0\\
     (a^2 + 1) | \tilde{\psi} (p) \rangle & =  0
\end{aligned} 
\quad&\rightarrow\quad
\begin{aligned}
     \partial^2 | \psi (x) \rangle & =  0\\
     (a^{\dag} \cdot \partial + i \mu) | \psi (x) \rangle & =  0\\
     (a \cdot \partial) | \psi (x) \rangle & =  0\\
     (a^2 + 1) | \psi (x) \rangle & =  0\quad . 
\end{aligned}
\end{align}
\section{Factorial power}\label{factorial}

We define the symbols known as factorial powers. It should be noted that each of the four symbols represents a product of $n$ factors. The definitions are as follows:
\begin{alignat}{3}
    (x)^{\underline{n}} & :=  x (x - 1) \cdots (x - n + 1) && , \quad  (x)^{\overline{n}} && :=  x (x + 1) \cdots (x + n - 1)\\
    (x)^{\underline{- n}} & :=  \frac{1}{(x + 1)^{\overline{n}}} && , \quad (x)^{\overline{- n}} && :=  \frac{1}{(x - 1)^{\underline{n}}} \qquad .
\end{alignat}
These factorial powers possess the properties of binomial polynomials:
\begin{align}
        (x + y)^{\underline{m}} &  =  \sum_{k = 0}^m \frac{m!}{k! (m - k) !} (x)^{\underline{k}} (y)^{\underline{m - k}}\\
      (x + y)^{\overline{m}} & =  \sum_{k = 0}^m \frac{m!}{k! (m - k) !} (x)^{\overline{k}} (y)^{\overline{m - k}}\quad .
\end{align}
These properties are used to demonstrate the existence of the inverse operator of the trace generating operator$Z_{X_k}^{\omega}$, as well as the properties associated with it, as described in Appendix \ref{traceop_inv}.

\section{Trace Generating Operator $Z_X^{\omega}$}\label{traceop}

\subsection{Coherent State Generating Operator $z^{\omega}$}\label{coherent}

Before diving into the main topic, let's briefly review the coherent state and its generating operator. Given creation and annihilation operators $b, b^{\dag}$, where $[b, b^{\dag}] = \hbar$, $b \hbar = \hbar b$, the vacuum state $|0 \rangle$ is introduced such that $b| 0 \rangle = 0$. Applying coherent state generating operator:
\begin{align}
    z^{\omega} &:= e^{\omega b^{\dag} / \hbar}
\end{align}
 to $|0 \rangle$ generates coherent state $| \omega \rangle$ indexed by $\omega$, which has the following properties:
\begin{align}
    | \omega \rangle  &= z^\omega | 0 \rangle \quad,&   (b - \omega)    z^\omega  &= z^\omega b\quad ,&  (b - \omega) | \omega \rangle  &=  0\quad ,& \langle \omega'  | \omega \rangle  &=  e^{\omega'^{\ast} \omega / \hbar^2}\quad .
\end{align}

\subsection{Trace Generating Operator $Z_X^{\omega}$}

Let's consider an operator $Z_X^{\omega}$ that, similar to the coherent state generating operator, acts on a traceless state $| \psi^0 \rangle $, given by $l_2 | \psi^0 \rangle = 0 $ to generate another state $| \psi^{\omega} \rangle $ with a fixed trace, given by $(l_2 + \omega) | \psi^{\omega} \rangle = 0 $. See {\cite{Najafizadeh:2017tin}} and {\cite{Alkalaev:2017hvj}} for similar operators. We assume that $Z_X^{\omega}$ can be expanded as a series in $l_2^{\dagger}$. From the algebraic relation $[l_2, l_2^{\dagger}] = X_1$, the expansion coefficients generally depend on $X$. We rewrite the condition $(l_2 + \omega) | \psi^{\omega} \rangle = 0$ as a difference equation for the expansion coefficients, and solve it to find $Z_X^{\omega}$. The result is as follows:
\begin{align}
  Z_X^{\omega} & :=  \sum_{m = 0}^{\infty} \frac{(- \omega l_2^{\dag})^m}{m!} (X)^{\underline{- m}} \quad .  \label{traceop_def_app}
\end{align}
Here, we have used the notation $(x)^{\underline{- n}}$, which represents a product of $n$ factors. Refer to Appendix \ref{factorial} for the definition. Using this form of $Z_X^{\omega}$, one can verify the following operator relation:
\begin{align}
  (l_2 + \omega) Z_X^{\omega} & =  Z_{X_2}^{\omega} l_2 \quad . \label{traceop_formula_app}
\end{align}
Thus for traceless state $ l_2 | \psi^0 \rangle  =  0$,
\begin{alignat}{2}
      (l_2 + \omega) | \psi^{\omega} \rangle & =  0 & , \quad | \psi^{\omega} \rangle & =  Z_X^{\omega} | \psi^0 \rangle \quad .
\end{alignat}
Here, we need to introduce a more general form of $Z_X^{\omega}$ by replacing $X$ with $X_k = X + k$:
\begin{align}
  Z_{X_k}^{\omega} & :=  \sum_{m = 0}^{\infty} \frac{(- \omega  l_2^{\dag})^m}{m!} (X_k)^{\underline{- m}} \quad . 
\end{align}

\subsubsection{Formulas between $Z_{X_k}^{\omega} $ and HS Operators, and the $k$-th Trace Generating Operator}

It can be directly verified through calculation that operator $Z_{X_k}^{\omega}$ possesses the following properties:
\begin{align}
  (l_2 + \omega)^k Z_{X_{1- k}}^{\omega} & =  Z_{X_{k + 1}}^{\omega} l_2^k  \label{traceop_formula1}\\
  (l_2 + \omega) Z_{X_k}^{\omega} & =  \omega Z_{X_{k + 1}}^{\omega}  \frac{k}{X_{k + 1}} + Z_{X_{k + 2}}^{\omega} l_2  \label{traceop_formula2}\\
  l_1 Z_{X_k}^{\omega} & =  Z_{X_{k + 1}}^{ \omega} l_1 + l_1^{\dag} Z_{X_{k + 1}}^{\omega} \frac{\omega}{X_{k + 1}} \quad . 
  \label{traceop_formula3}
\end{align}
Here, $ X_k  =  X + k  , \forall k \in \mathbb{Z}$.

We can leverage property (\ref{traceop_formula1}) to demonstrate that ``$k$-th trace generating operator'' $Z_{X_{1- k}}^{\omega}$ acts on $k$-th traceless state $| \psi_k^0 \rangle$ (where $l_2^k | \psi_k^0 \rangle = 0$) to create a new state $| \psi_k^{\omega} \rangle$, satisfying the condition $(l_2 + \omega)^k | \psi_k^{\omega} \rangle = 0 $:
\begin{alignat}{3}
    (l_2 + \omega)^k | \psi_k^{\omega} \rangle & =  0 & , \quad | \psi_k^{\omega}  \rangle & =  Z_{X_{1- k}}^{\omega} | \psi_k^0 \rangle & , \quad l_2^k |    \psi_k^0 \rangle & =  0 .
\label{traceop_prop}
\end{alignat}
\subsubsection{Inverse Operator of $Z_{X_k}^{\omega}$ and Useful Relations}\label{traceop_inv}

Here, the proof of the eqs.(\ref{traceop_inv_def}) and (\ref{traceop_inv_formula}) below requires the use of the properties of binomial polynomials associated with factorial powers in Appendix \ref{factorial}.

The inverse operator of $Z_{X_k}^{\omega}$ is found to exist as follows:
\begin{alignat}{2}
    (Z_{X_k}^{\omega})^{- 1} & =  \sum_{m = 0}^{\infty} (X_k)^{\overline{-m}} \frac{(\omega l_2^{\dag})^m}{m!} & , \quad & Z_{X_k}^{\omega} (Z_{X_k}^{\omega})^{- 1}  =  (Z_{X_k}^{\omega})^{- 1} Z_{X_k}^{\omega}   =  1 \quad .
\label{traceop_inv_def}
\end{alignat}

This reveals that any two $k$-th trace constrained states with $\omega$ and $\omega'$ are related as follows:
\begin{align}
    | \psi_k^{\omega'} \rangle & =  Z_{X_{- k + 1}}^{\omega'} (Z_{X_{- k +1}}^{\omega})^{- 1} | \psi_k^{\omega} \rangle \quad .
\end{align}
Furthermore, through somewhat cumbersome calculations, the following useful relations can be confirmed:
\begin{align}
  (Z_{X_{k - 1}}^{\omega})^{- 1} Z_{X_k}^{\omega} & =  1 + l_2^{\dag}  \frac{\omega}{X_k X_{k + 1}} \quad .  
\label{traceop_inv_formula}
\end{align}

\subsubsection{Normalization}

Consider the normalization of $| \psi_k^{\omega} \rangle$. Define
\begin{align}
    g_{\omega, k} & :=  \langle \psi_k^0 | Z_{X_k}^{\omega \dag} Z_{X_k}^{\omega} | \psi_k^0 \rangle 
\end{align}
and
\begin{align}
    \overline{Z}_{X_k}^{\omega} & :=  \frac{1}{\sqrt{g_{\omega, k}}}    Z_{X_k}^{\omega} \quad,
\end{align}
then
\begin{align}
    |\overline{ \psi}_k^{\omega} \rangle & :=  \overline{Z}_{X - k + 1}^{\omega} | \psi_k^0 \rangle
\end{align}
is normalized:
\begin{align}
    \langle \overline{\psi}_k^{\omega} | \overline{\psi}_k^{\omega} \rangle & =  1\quad .
\end{align}
\section{Derivation of Modified Wigner Equations using HS Operators}\label{m_wigner}

\subsection{Rewriting Using the Trace Generating Operator $Z_X^{\omega}$}

Starting with the original Wigner eqs.(\ref{kg_hs})-(\ref{trace_hs}) expressed in terms of HS operators:
\begin{align}
  \partial^2 | \psi^{\omega} \rangle & =  0   \label{kg_hs_app}\\
  (l_1^{\dag} - i \mu) | \psi^{\omega} \rangle & =  0  \label{gauge_hs_app}\\
  l_1 | \psi^{\omega} \rangle & =  0   \label{div_hs_app}\\
  (l_2 + \omega) | \psi^{\omega} \rangle & =  0 \quad,   \label{trace_hs_app}
\end{align}
we substitute $| \psi^{\omega} \rangle  =  Z_X^{\omega} | \psi^0 \rangle$ into eqs.(\ref{kg_hs_app})-(\ref{trace_hs_app}), where $Z_X^{\omega}$ is trace generating operator (\ref{traceop_def_app}). While eq.(\ref{kg_hs_app}) trivially gives $\partial^2 | \psi^0 \rangle = 0$, eq.(\ref{trace_hs_app}) is rewritten by using the defining equation (\ref{traceop_formula_app}) of $Z_X^{\omega}$ as:
\begin{gather}
    0=(l_2 + \omega) | \psi^{\omega} \rangle  =  (l_2 + \omega) Z_X^{\omega} | \psi^0 \rangle  =  Z_{X_2}^{\omega} l_2 | \psi^0 \rangle\quad .
\end{gather}
Therefore, by multiplying by $(Z_{X_2}^{\omega})^{- 1}$,\footnote{The inverses of operators $Z$s are explicitly written; refer to Appendix \ref{traceop_inv}.
}
we obtain:
\begin{align}
  l_2 | \psi^0 \rangle & =  0 \quad . 
\end{align}
The LHS of eq.(\ref{gauge_hs_app}) is calculated as follows:
\begin{align}
  (l_1^{\dag} - i \mu) | \psi^{\omega} \rangle & =  (l_1^{\dag} - i \mu)
  Z_X^{\omega} | \psi^0 \rangle 
\notag\\
  & =  (Z_{X_{- 1}}^{\omega} l_1^{\dag} - i \mu Z_X^{\omega}) | \psi^0
  \rangle 
\notag\\
  & =  Z_{X_{- 1}}^{\omega} \{ l_1^{\dag} - i \mu (Z_{X_{- 1}}^{\omega})^{-
  1} Z_X^{\omega} \} | \psi^0 \rangle \quad . 
\end{align}
Using formula (\ref{traceop_inv_formula}) with $k = 1$: $(Z_{X_{- 1}}^{\omega})^{- 1} Z_X^{\omega} =  1 + l_2^{\dag}\frac{\omega}{X X_1}$, and multiplying by $(Z_{X_{- 1}}^{\omega})^{- 1}$, we get:
\begin{align}
  \left( l_1^{\dag} - i \mu - i \mu l_2^{\dag} \frac{\omega}{X X_1} \right) |
  \psi^0 \rangle & =  0 \quad .  \label{gauge_mod_app00}
\end{align}
Finally, by using the formula (\ref{traceop_formula3}), the LHS of eq.(\ref{div_hs_app}) is calculated as follows. First, we eliminate the $l_1^{\dag}$ term using eq.(\ref{gauge_mod_app00}) and then use eq.(\ref{traceop_inv_formula}) again:
\begin{align}
  l_1 | \psi^{\omega} \rangle & =  l_1 Z_X^{\omega} | \psi^0 \rangle
\notag\\
  & =  \left( Z_{X_1}^{\omega} l_1 + Z_X^{\omega} \frac{\omega}{X}  l_1^{\dag} \right) | \psi^0 \rangle 
\notag\\
  & =  \left\{ Z_{X_1}^{\omega} l_1 + i \mu Z_X^{\omega} \frac{\omega}{X}  \left( 1 + l_2^{\dag} \frac{\omega}{X X_1} \right) \right\} | \psi^0 \rangle
\notag\\
  & =  \left\{ Z_{X_1}^{\omega} l_1 + i \mu Z_X^{\omega} \left( 1 +  l_2^{\dag} \frac{\omega}{X_1 X_2} \right) \frac{\omega}{X} \right\} | \psi^0  \rangle 
\notag\\
  & =  Z_{X_1}^{\omega} \left( l_1 + i \mu \frac{\omega}{X} \right) | \psi^0 \rangle \quad . 
\end{align}
Therefore, by multiplying by $(Z_{X_1}^{\omega})^{- 1}$, eq.(\ref{div_hs_app}) gives:
\begin{align}
  \left( l_1 + i \mu \frac{\omega}{X} \right) | \psi^0 \rangle & =  0 \quad .
\end{align}
Summarizing, we have the following modified Wigner equations for the state $|\psi^0 \rangle$:
\begin{align}
  \partial^2 | \psi^0 \rangle & =  0  
\label{kg_mod_app0}\\
  \left( l_1^{\dag} - i \mu - i \mu l_2^{\dag} \frac{\omega}{X X_1} \right) |
  \psi^0 \rangle & =  0  
\label{gauge_mod_app0}\\
  \left( l_1 + i \mu \frac{\omega}{X} \right) | \psi^0 \rangle & =  0 
\label{div_mod_app0}\\
  l_2 | \psi^0 \rangle & =  0 \quad .  
\label{trace_mod_app0}
\end{align}

\subsection{Phase Rotation for Real field formulation}

To investigate if state vectors with real fields can satisfy Wigner equations, we'll explore rewriting Wigner equations without involving imaginary units. To do this, we'll redefine the state vector using a phase factor dependent on $N$. Let's rewrite the phase rotation as follows and find $\varepsilon_N$:
\begin{align}
    | \psi^0 \rangle & =  e^{i \varepsilon_N} | \Phi \rangle \quad .
\end{align}
Then, \ eqs.(\ref{gauge_mod_app0}) and (\ref{div_mod_app0}) can be rewritten as follows:
\begin{align}
    \left( l_1^{\dag} - i \mu - i \mu l_2^{\dag} \frac{\omega}{X X_1} \right)    | \psi^0 \rangle & =  e^{i \varepsilon_{N - 1}} \left(l_1^{\dag} - i    e^{i (\varepsilon_N - \varepsilon_{N - 1})} \mu - i e^{i (\varepsilon_{N - 2} - \varepsilon_{N - 1})} \mu l_2^{\dag} \frac{\omega}{X X_1} \right) |\Phi \rangle
\\
    \left( l_1 + i \mu \frac{\omega}{X} \right) | \psi \rangle & =  e^{i  \varepsilon_{N + 1}} \left( l_1 + i e^{i (\varepsilon_N - \varepsilon_{N +
    1})} \mu \frac{\omega}{X} \right) | \Phi \rangle \quad.
\end{align}
The parentheses on the right-hand side contain the phases \ $i e^{i(\varepsilon_N - \varepsilon_{N - 1})}$, $i e^{i (\varepsilon_N - 2 -\varepsilon_{N - 1})}$, and $i e^{i (\varepsilon_N - \varepsilon_{N + 1})}$. Let's consider if we can determine $\varepsilon_N$ to make all of them real. For example, we may choose:
\begin{align}
  \varepsilon_n - \varepsilon_{n - 1} & =  - \pi /2 \text{ for all } n \in \mathbb{Z} \quad . 
\end{align}
This is achieved when:
\begin{align}
  \varepsilon_n & =  - \frac{\pi}{2} n \quad . 
\end{align}
Thus, by phase rotating as
\begin{align}
  | \psi^0 \rangle & =  e^{- i \frac{\pi}{2} N} | \Phi \rangle \quad ,
\end{align}
Wigner equations are rewritten without imaginary factors as:
\begin{align}
  \partial^2 | \Phi \rangle & =  0  \label{kg_mod_app}
\\
  \left( l_1^{\dag} - \mu + \mu l_2^{\dag} \frac{\omega}{X X_1} \right) | \Phi
  \rangle & =  0  \label{gauge_mod_app}
\\
  \left( l_1 - \mu \frac{\omega}{X} \right) | \Phi \rangle & =  0 
  \label{div_mod_app}
\\
  l_2 | \Phi \rangle & =  0 \quad .  \label{trace_mod_app}
\end{align}

\section{Quadratic Casimir Operator $C_2$}\label{casimir}

We will demonstrate that the quadratic Casimir operator $C_2$ satisfies the Casimir condition $  C_2 | \Phi \rangle = \mu^2 | \Phi \rangle$ using modified Wigner eqs.(\ref{kg_mod})-(\ref{trace_mod}) through direct calculation. $C_2$ is expressed as:\footnote{Please refer to equations (6) and (7) in {\cite{Bargmann:1948ck}}.
}
\begin{alignat}{3}
    C_2 & =  \frac{1}{6} v^{\mu \nu \lambda} v_{\mu \nu \lambda} & ,  \quad v_{\nu
    \rho \sigma} & =  \partial_{\nu} S_{\rho \sigma} + \text{cyclic} & , \quad
    S_{\mu \nu} & =  i (a^{\dag}_{\mu} a_{\nu} - a^{\dag}_{\nu} a_{\mu}) \quad .
\end{alignat}
We have:
\begin{align}
  C_2 & =  - (\partial^{\mu} S _{\mu \lambda}) (\partial_{\nu} S^{\nu \lambda}) + \frac{1}{2} \partial^2 S^{\nu \lambda} S_{\nu \lambda} 
\label{Casimir}\\
  \partial_{\nu} S^{\nu \lambda} & =  i \left( l_1^{\dag} a^{\lambda} +  a^{\dag \lambda} {l_1}  \right) \quad . 
\end{align}
Firstly, using the masslessness condition  (\ref{kg_mod}), the second term of eq. (\ref{Casimir}) vanishes on $|\Phi \rangle$. Next, evaluating the first term of eq. (\ref{Casimir}) on $| \Phi \rangle$ through direct calculation using the divergence condition  (\ref{div_mod}), we find:
\begin{align}
  C_2 | \Phi \rangle & =  2 \mu \omega \left( l_1^{\dag} + l_2^{\dag}  \frac{\mu \omega}{{X X_1} } \right) | \Phi \rangle \quad . 
\end{align}
Finally, applying eq.(\ref{gauge_mod}), one arrives at:
\begin{align}
    C_2 | \Phi \rangle & =  2 \mu^2 \omega | \Phi \rangle \quad .
\end{align}
By setting $\omega = 1 / 2$, this correctly yields the Casimir eigenvalue on the state vector for CS:
\begin{align}
  C_2 | \Phi \rangle & =  \mu^2 | \Phi \rangle \quad . 
\end{align}

\section{Algebraic Structure of Wigner Equations}\label{alg_wigner}

We will investigate an algebraic structure of modified Wigner eqs.(\ref{kg_mod})-(\ref{trace_mod}) to understand their interrelationships. To do this, we define factors acting on the state vector $|\Phi \rangle$ as follows:
\begin{align}
  T_1 & :=  \partial^2 \\
  T_2 & :=  l_1^{\dag} - \mu + \mu l_2^{\dag} \frac{\omega}{X X_1} \\
  T_3 & :=  l_1 - \mu \frac{\omega}{X} \\
  T_4 & :=  l_2 \quad . 
\end{align}
Then, modified Wigner eqs.(\ref{kg_mod})-(\ref{trace_mod}) and eq.(\ref{gauge_invsol}) can be expressed as:
\begin{align}
  T_i | \Phi \rangle & =  0, \; i = 1 \sim 4 \\
  | \Phi \rangle  & =  \delta (T_2) | \tilde{\phi} \rangle \quad . 
\end{align}
Direct calculations confirm that the operators $T_i$ have the following commutation relations:
\begin{align}
  T_1 T_i & =  T_i T_1,\; i = 2, 3, 4  \label{Walg1i}\\
  T_3 T_2 & =  T_2' T_3 - T_1  \label{Walg32}\\
  T_4 T_2 & =  T_2'' T_4 - T_3  \label{Walg42}\\
  T_4 T_3 & =  T_3' T_4 \quad .  \label{Walg43}
\end{align}
Here,
\begin{alignat}{3}
    T_2' & :=  l_1^{\dag} - \mu + \mu l_2^{\dag} \frac{\omega}{X_1 X_2} & , \quad  T_2'' & :=  l_1^{\dag} - \mu + \mu l_2^{\dag} \frac{\omega}{X_2 X_3} & ,\quad  T_3' & :=  l_1 - \mu \frac{ \omega}{X_2}\quad.
\end{alignat}

\subsection{Non-Independence of Wigner Equations}

Eq.(\ref{Walg42}) shows that the equation for $T_3$ can be derived from those for $T_2$ and $T_4$. Similarly, eq.(\ref{Walg32}) shows that the equation for $T_1$ can be derived from those for $T_2$ and $T_3$:
\begin{align}
  T_3 | \Phi \rangle  & =  T_2'' T_4 | \Phi \rangle - T_4 T_2 | \Phi \rangle = 0   \\
  T_1 | \Phi \rangle  & =  T_2' T_3 | \Phi \rangle - T_3 T_2 | \Phi \rangle =  0 \quad . 
\end{align}

\subsection{Condition on $| \Phi \rangle$ induced by conditions on $|\tilde{\phi} \rangle$}

Eq.(\ref{Walg1i}) for $i = 2$, along with eqs.(\ref{Walg32}) and (\ref{Walg42}), implies that if $T_{1, 3, 4}$ vanish on $| \tilde{\phi} \rangle $, then they also vanish on $T_2 | \tilde{\phi} \rangle $ and consequently on $| \Phi \rangle = \delta (T_2) | \tilde{\phi} \rangle  $. Therefore, together with the delta function property $T_2 \delta (T_2) | \tilde{\phi} \rangle = 0, $ we have shown that:
\begin{align}
	\text{if } T_{1, 3, 4} | \tilde{\phi} \rangle   =  0 , \quad &\text{ then } T_{1\sim 4} |\Phi \rangle   =  0 \text{ where } | \Phi \rangle   =  \delta (T_2) | \tilde{\phi} \rangle\quad.
\end{align}

\section{Introduction Method of Compensator Fields to Remove Constraints on Gauge Parameter}\label{compensator}

In this appendix, we aim to find a gauge invariant Lagrangian under unconstrained gauge parameter.

In Subsection \ref{compensator_general}, we examine a simple general model that emphasizes the structure of the trace operator $l_2$ and the gauge parameter $\lambda$ in the Lagrangian $\mathcal{L}_4$(\ref{Lag_4f}). Subsequently, we explore in the general model how to introduce a compensator field into Lagrangian to maintain gauge invariance even under unconstrained gauge parameter, thus avoiding trace constraints.

In Subsection \ref{compensator_apply}, we apply this method to the Lagrangian $\mathcal{L}_4$(\ref{Lag_4f}) and derive the Lagrangian
$\mathcal{L}_5$(\ref{Lag_5f}).

\subsection{General Method for Relaxing the Traceless Constraint for the Gauge Parameter}\label{compensator_general}

Suppose we have a Lagrangian $\mathcal{L}$ with unconstrained several fields $\phi_i$ as follows(repeated $i, j$ are summed):
\begin{align}
  \mathcal{L} & =  \langle \phi_i |L_{i j} | \phi_j \rangle \quad . 
\end{align}
This varies under transformation with unconstrained gauge parameter $\lambda$, and assume it vanish if $\lambda$ would be traceless $l_2 | \lambda \rangle = 0$. This must be written as follows:\footnote{We can consider variation for field in 'Ket' state only without loss of generality.
}
\begin{align}
  \delta | \phi_j \rangle & =  P_j | \lambda \rangle \nonumber
\\
  \delta \mathcal{L} & =  \langle \phi_i |L_{i j} P_j | \lambda \rangle =  \langle \phi_i |O_i l_2 | \lambda \rangle, \text{i.e.,} L_{i j} P_j \equiv
  O_i l_2  \quad . 
\end{align}
Here, $L_{i j}$, $O_i$, $P_i$ are given operators. We add a new field $\psi$ whose gauge transformation is
\begin{align}
  \delta | \psi \rangle  & =  l_2 | \lambda \rangle  
\end{align}
and new Lagrangian $\widetilde{\mathcal{L}}$ are introduced with $\psi$ as:
\begin{align}
    \widetilde{\mathcal{L}} & :=  \mathcal{L} + \langle \phi_i |L_{i \psi} |    \psi \rangle + \langle \psi |L_{i \psi}^{\dag} | \phi_i \rangle + \langle  \psi |L_{\psi \psi} | \psi \rangle  \quad .
\label{Lag_new}
\end{align}
Here, $L_{i \psi}$ and $L_{\psi \psi}$ are unknown operators those will be determined by requiring gauge invariance under unconstrained gauge parameter $\lambda$. Variation of $\widetilde{\mathcal{L}}$ under transformation of $\phi_i$ and $\psi$ is calculated as:
\begin{align}
  \delta \widetilde{\mathcal{L}} & =  \langle \phi_i | (O_i + L_{i \psi})   l_2 | \lambda \rangle + \langle \psi | (L_{i \psi}^{\dag} P_i +
  L_{\psi \psi} l_2) | \lambda \rangle \quad . 
\end{align}
Gauge invariance for unconstrained fields and the gauge parameter requires the following conditions:
\begin{align}
  O_i + L_{i \psi} & =  0 \\
  L_{i \psi}^{\dag} P_i + L_{\psi \psi} l_2 & =  0 \quad . 
\end{align}
These equations determine $L_{i \psi}$ \ and $L_{\psi \psi}$:
\begin{align}
  L_{i \psi} & =  - O_i \\
  L_{\psi \psi} l_2 & =  - L_{i \psi}^{\dag} P_i = O^{\dag}_i P_i \quad . 
\end{align}
Note that it is necessary for $O_i^{\dagger} P_i$ to have a form like $(\ldots) l_2$ in order to find $L_{\psi \psi}$ from the above relation.

\subsection{Applying the General Method to Derive $\mathcal{L}_5$(\ref{Lag_5f}) from $\mathcal{L}_4$(\ref{Lag_4f})}\label{compensator_apply}

We will now apply the general method described in Subsection \ref{compensator_general} to find Lagrangian with an unconstrained gauge parameter for our Lagrangian $\mathcal{L}_4$(\ref{Lag_4f}). The fields used in Subsection \ref{compensator_general} correspond to the following in $\mathcal{L}_4$:
\begin{align}
  \langle \phi_i | & =  
  \begin{pmatrix}
        \langle s_1 | & \langle s_2 |& \langle a_2 | & \langle a_3 |
    \end{pmatrix}, \; \psi = s_4, \;\lambda  = \lambda 
\end{align}
And the gauge transformation and the variation of the Lagrangian $\mathcal{L}_4$ under it can be written as follows:
\begin{alignat}{2}
    \delta | \phi_j \rangle  &=  P_j | \lambda \rangle & ,\quad  
    \delta \mathcal{L}_4  =  \langle \phi_i |L_{ij} P_j | \lambda \rangle &= 
    \begin{pmatrix}
      \langle s_1 | & \langle s_2 | & \langle a_2 | & \langle a_3 |
    \end{pmatrix} 
    \begin{pmatrix}
 	     0\\
         0\\
         \tilde{L}_{X_2}^{\dag}\\
         - L'_{X_2}
   \end{pmatrix}  l_2 | \lambda \rangle  
\notag \\
&&& \equiv  \langle \phi_i |O_i l_2 | \lambda \rangle 
   \quad .
\end{alignat}
From these we can find $O$ and $P$ as:
\begin{alignat}{2}
    P & =  
    \begin{pmatrix}
      L_X^{\dag} + l^{\dag}_2 \frac{\mu}{X \sqrt{2 X_1}}
\\
      \sqrt{X_1} L_X \frac{1}{\sqrt{X}}
\\
      - (\partial^2_{} + L_{X_2}^{\dag} L_{X_2})  \frac{\mu}{X \sqrt{2 X_1}}
\\
      - L_{X_3} L_{X_2}  \frac{\mu}{X \sqrt{2 X_1}}
    \end{pmatrix} & ,  \quad 
     O  & =  
     \begin{pmatrix}
      0\\
      0\\
      \tilde{L}_{X_2}^{\dag}\\
      - L'_{X_2}    
    \end{pmatrix} \quad .\label{OP}
\end{alignat}
From the relation $L_{\psi} = - O$, $L_{\psi}$ is determined as:
\begin{align}
  L_{\psi} & = 
  \begin{pmatrix}
    0\\
    0\\
    - \tilde{L}_{X_2}^{\dag}\\
    L'_{X_2}
   \end{pmatrix} \quad . 
\end{align}
From eq.(\ref{OP}), we calculate $O^{\dag}_i P_i$ for each order of $\mu$ and then summarize it as:
\begin{align}
  O^{\dag}_i P_i & =  - \frac{\mu}{X_2  \sqrt{2 X_3}}  (\partial^2 +  L_{X_4}^{\dag} L_{X_4})  \frac{\mu}{X_2  \sqrt{2 X_3}} l_2 \quad . 
\end{align}
Therefore, from the relation \ $  L_{\psi \psi} l_2  =  O^{\dag}_i P_i $, we can determine \ $L_{\psi \psi}$ as:
\begin{align}
  L_{\psi \psi} & =  - \frac{\mu}{X_2  \sqrt{2 X_3}}  (\partial^2 +  L_{X_4}^{\dag} L_{X_4})  \frac{\mu}{X_2  \sqrt{2 X_3}} \quad . 
\end{align}
This yields the Lagrangian $\mathcal{L}_5$(\ref{Lag_5f}), which is gauge invariant under the trace-full gauge parameter $\lambda$, as shown by the construction method above.

\section{Equivalence of Unconstrained Lagrangian $\mathcal{L}_5$ (\ref{Lag_5f}) \\and Fronsdal-like Lagrangian $\mathcal{L}_1$ (\ref{Lag_1f})} \label{eom_equiv}

First, let's consider the equivalence between $\mathcal{L}_5$(\ref{Lag_5f}) and $\mathcal{L}_4$(\ref{Lag_4f}).  This can be clarified as follows. Utilizing the gauge transformation:
\begin{align}
  \delta |s_4 \rangle &= l_2 | \lambda \rangle,
\end{align}
we gauge-fix $s_4$ to zero. As a consequence, the gauge parameter acquires constraint $l_2 | \lambda \rangle = 0$, leading to the reduction of the Lagrangian to $\mathcal{L}_4$.

Now, to establish the equivalence between $\mathcal{L}_4$ and $\mathcal{L}_1$(\ref{Lag_1f}), we must examine the equality of their EOM. The detailed proof is presented below.

\subsection{Derivation of EOM for Fronsdal-like Lagrangian $\mathcal{L}_1$(\ref{Lag_1f})}\label{eom_frions}

Here, we derive EOM for the Lagrangian $\mathcal{L}_1$(\ref{Lag_1f}). By varying it with respect to $\langle \phi | $ under the condition $l_2^2 | \phi \rangle = 0$, the EOM can be expressed as:
\begin{align}
  P_2 E | \phi \rangle & =  0 \quad . 
\end{align}
Here, $E$ is defined as follows(see also Appendix \ref{convention}):
\begin{align}
  	E & :=\begin{pmatrix}
	1 & l^{\dag}_2
\end{pmatrix} 
\begin{pmatrix*}[c]
	\partial^2  + L_X^{\dag} L_X & L_X^{\dag} L_{X_1}^{\dag}
\\
	L_{X_1} L_X & - 2 \partial^2  + L_{X_2}^{\dag} L_{X_2}
\end{pmatrix*} 
\begin{pmatrix}
	1\\
	l_2
\end{pmatrix}
\label{EOMop_def}
\end{align}
and $P_2$ is a projection operator onto the double-traceless field, satisfying $P_2 | \phi \rangle = | \phi \rangle$.\footnote{See Appendix \ref{convention} for the projection operator $P_2$.
}
By direct calculation, one finds that $E | \phi \rangle$ is triple traceless:
\begin{align}
  l_2^3 E | \phi \rangle & =  0 \quad . 
\end{align}
For such a state, $P_2$ can be expressed simply as:
\begin{align}
  P_2 E | \phi \rangle & =  \Pi_2 E | \phi \rangle \quad . 
\end{align}
Here,
\begin{align}
    \Pi_2 & :=  1 - l_2^{\dag 2}  \frac{1}{2 X_1 X_2} l_2^2
\end{align}
is a projection operator onto double-traceless fields within triple traceless space(see eq.(\ref{proj}) in Appendix \ref{convention}). The calculation of $\Pi_2 E$ yields:
\begin{align}
  \Pi_2 E & =  E + l_2^{\dag 2} L_{X_3}  \frac{\mu}{X_1  \sqrt{2 X_2}} l_2 +  (\ldots) l_2^2 \quad . 
\end{align}
Thus, EOM $P_2 E | \phi \rangle = 0$ is expressed as:
\begin{align}
  \left( E + l_2^{\dag 2} L_{X_3}  \frac{\mu}{X_1  \sqrt{2 X_2}} l_2 \right) |  \phi \rangle & =  0 \quad .  
\label{EOM_single}
\end{align}

In Subsection \ref{eom_unconstl}, starting from the Lagrangian $\mathcal{L}_4$(\ref{Lag_4f}), we derive this equation along with the double-traceless property $l_2^2 | \phi \rangle = 0$. It is noteworthy that, irrespective of the EOM itself, as evident from $l_2^2 P_2 E | \phi \rangle = 0$, the left-hand side of eq.(\ref{EOM_single}) is already double-traceless:
\begin{align}
  l_2^2 \left( E + l_2^{\dag 2} L_{X_3}  \frac{\mu}{X_1  \sqrt{2 X_2}} l_2  \right) | \phi \rangle & =  0 \quad .  
  \label{EOM_dtrace}
\end{align}

\subsection{Derivation of EOM of $\mathcal{L}_4$(\ref{Lag_4f}) and its Identity with EOM of $\mathcal{L}_1 $(\ref{Lag_1f})}\label{eom_unconstl}

The equivalence of the EOM between $\mathcal{L}_4$(\ref{Lag_4f}) and $\mathcal{L}_1$(\ref{Lag_1f}) can be proved as follows. To begin, the EOM of $\mathcal{L}_4$ consist of four equations:
\begin{align}
\begin{pmatrix}
	\partial^2_{} + L_X^{\dag} L_X & - L_X^{\dag} L_{X_1}^{\dag} &   l_2^{\dagger} & 0
\\
    - L_{X_1} L_X & - 2 \partial^2_{} + L_{X_2}^{\dag} L_{X_2} & 1 & - l_2^{\dag}
\\
    l_2 & 1 & 0 & 0
\\
    0 & - l_2 & 0 & 0
\end{pmatrix} 
\begin{pmatrix}
    | s_1 \rangle\\
    | s_2 \rangle\\
    | a_2 \rangle\\
    | a_3 \rangle
\end{pmatrix} & =  0 \quad .  
\label{EOM_4f}
\end{align}
Using the EOM of $\langle a_3 |$, $|s_2 \rangle$ is tracelessly constrained:
\begin{align}
  l_2 |s_2 \rangle & =  0 \quad . 
\end{align}
Subsequently, using the EOM of $\langle a_2 |$, $|s_2 \rangle$ can be expressed as:
\begin{align}
  |s_2 \rangle &= - l_2 |s_1 \rangle \quad,
\end{align}
imposing double-traceless constraint on $|s_1 \rangle$ from tracelessness of $|s_2 \rangle$. The remaining two equations involve $| a_2 \rangle$, $| a_3 \rangle$ and $| s_1 \rangle$, where \ $| s_1 \rangle$ is constrained to be double-traceless:
\begin{align}
  \begin{pmatrix}
    \partial^2_{} + L_X^{\dag} L_X & - L_X^{\dag} L_{X_1}^{\dag}
    \\
    - L_{X_1} L_X & - 2 \partial^2_{} + L_{X_2}^{\dag} L_{X_2}
  \end{pmatrix} 
  \begin{pmatrix}    
  	 1
  	 \\
    - l_2
  \end{pmatrix}
     | s_1 \rangle 
  + 
  \begin{pmatrix}  	l_2^{\dagger} & 0
    \\
    1 & - l_2^{\dag}
  \end{pmatrix} 
  \begin{pmatrix}    | a_2 \rangle
    \\
    | a_3 \rangle
  \end{pmatrix}  
  & =  0  
\label{EOM_3f}\\
  l_2^2 | s_1 \rangle 
  & =  0 \quad . 
\end{align}
The lower components of eq.(\ref{EOM_3f}) yield the solution for $| a_2 \rangle$ as:
\begin{align}
  | a_2 \rangle & =  l_2^{\dag} | a_3 \rangle 
  + \begin{pmatrix}
    L_{X_1} L_X, & 2 \partial^2_{} - L_{X_2}^{\dag} L_{X_2}
    \end{pmatrix} 
    \begin{pmatrix}
    1\\
    - l_2
    \end{pmatrix}
    | s_1 \rangle \quad .  
    \label{a2_sol}
\end{align}
By multiplying eq.(\ref{EOM_3f}) by $\begin{pmatrix} 1 & - l_2^{\dag}\end{pmatrix}$ from the left and using definition (\ref{EOMop_def}) of $E$, one obtains:
\begin{alignat}{2}
    E | s_1 \rangle + l_2^{\dag 2} | a_3 \rangle & =  0 & , \quad l_2^2 | s_1 \rangle &= 0 \quad .  
    \label{EOM_2f}
\end{alignat}
By multiplying eq.(\ref{EOM_2f}) by $l_2^2$ and using property (\ref{EOM_dtrace}), we obtain:
\begin{align}
  l_2^2 l_2^{\dag 2} \left( | a_3 \rangle - L_{X_3}  \frac{\mu}{X_1  \sqrt{2 X_2}} l_2 | s_1 \rangle \right) & =  0 \quad . 
\end{align}
Since it can be proved that any state $| \ldots \rangle$ satisfying $ l_2^2 l_2^{\dag 2} | \ldots \rangle = 0$ is zero: $| \ldots . \rangle = 0 $, $| a_3 \rangle$ is solved as:
\begin{align}
  | a_3 \rangle & =  L_{X_3}  \frac{\mu}{X_1  \sqrt{2 X_2}} l_2 | s_1 \rangle \quad . 
\end{align}
By substituting this into eq.(\ref{EOM_2f}), we obtain:
\begin{alignat}{2}
    \left( E + l_2^{\dag 2} L_{X_3}  \frac{\mu}{X_1  \sqrt{2 X_2}} l_2 \right) | s_1 \rangle & =  0 & , \quad 
    l_2^2 | s_1 \rangle &= 0 \quad .  
    \label{EOM_1f}
\end{alignat}
This aligns with the EOM (\ref{EOM_single}) by $\phi \equiv s_1$. Starting from the EOM (\ref{EOM_4f}), we derived both the EOM (\ref{EOM_single}) and the double-traceless constraint $l_2^2 | \phi \rangle = 0$, establishing their equivalence. This proves equivalence between the Lagrangians $\mathcal{L}_4$(\ref{Lag_4f}) and $\mathcal{L}_1$(\ref{Lag_1f}).

\medskip
\printbibliography[heading=bibintoc,title={References}]
\end{document}